\documentclass[acmsmall]{acmart}

\setcopyright{acmcopyright}
\acmJournal{CSUR}
\acmYear{2024} \acmVolume{1} \acmNumber{1} \acmArticle{1} \acmMonth{1} \acmPrice{XXX}\acmDOI{XXX/XXXX}

\usepackage{amsmath, amsfonts}
\usepackage{algorithmic, algorithm}
\usepackage{subfigure}
\usepackage{graphicx}
\usepackage{textcomp}
\usepackage{xcolor}
\usepackage{upgreek} 
\usepackage{threeparttable}
\usepackage{booktabs,tabularx,dcolumn}
\usepackage{multirow}
\usepackage{pifont}
\usepackage[misc]{ifsym}

\graphicspath{ {./figures/}{./photos/} }

\title{A Survey on UAV-enabled Edge Computing: Resource Management Perspective}

\author{Xiaoyu Xia}
\orcid{0000-0003-3526-3217}
\affiliation{%
  \institution{School of Computing Technologies, RMIT University}
  \state{VIC}
  \postcode{3000}
  \country{Australia}}
\email{xiaoyu.xia@rmit.edu.au}

\author{Sheik Mohammad Mostakim Fattah}
\affiliation{%
  \institution{Centre for Research on Engineering Software Technologies, University of Adelaide}
  \state{SA}
  \postcode{5005}
  \country{Australia}}
\email{sheik.fattah@adelaide.edu.au}

\author{Muhammad Ali Babar}
\affiliation{%
  \institution{Centre for Research on Engineering Software Technologies, University of Adelaide}
  \state{SA}
  \postcode{5005}
  \country{Australia}}
\email{ali.babar@adelaide.edu.au}

\begin{document}

\begin{abstract}
Edge computing facilitates low-latency services at the network's edge by distributing computation, communication, and storage resources within the geographic proximity of mobile and Internet-of-Things (IoT) devices. The recent advancement in Unmanned Aerial Vehicles (UAV) technologies has opened new opportunities for edge computing in military operations, disaster response, or remote areas where traditional terrestrial networks are limited or unavailable. In such environments, UAVs can be deployed as aerial edge servers or relays to facilitate edge computing services. This form of computing is also known as UAV-enabled Edge Computing (UEC), which offers several unique benefits such as mobility, line-of-sight, flexibility, computational capability, and cost-efficiency. However, the resources on UAVs, edge servers, and IoT devices are typically very limited in the context of UEC. Efficient resource management is, therefore, a critical research challenge in UEC. In this article, we present a survey on the existing research in UEC from the resource management perspective. We identify a conceptual architecture, different types of collaborations, wireless communication models, research directions, key techniques and performance indicators for resource management in UEC. We also present a taxonomy of resource management in UEC. Finally, we identify and discuss some open research challenges that can stimulate future research directions for resource management in UEC.
\end{abstract}



\maketitle

\section{Introduction}

\subsection{Edge Computing and UAVs}

{
The recent advancements of the Internet of Things (IoT) and wireless communication technologies have introduced many new applications that require high computational power and low latency \cite{mao2017survey}, including wearable cognitive assistance, augmented reality (AR), smart healthcare, facial recognition, and road safety monitoring \cite{zhao2021fairness}. However, IoT devices typically have limited computational resources, storage, network coverage, and energy. Therefore, resource-intensive IoT applications often face significant challenges in maintaining the expected Quality of Services (QoS) \cite{mach2017mobile, li2022budget}. A conventional approach to maintain the expected QoS is to leverage cloud computing \cite{lin2020survey} by utilizing resources from remote cloud servers in different forms such as virtual machines, virtual storage, and virtual private networks \cite{armbrust2010view}.
However, cloud computing is now considered inadequate to meet the low-latency requirements of resource-intensive and delay-sensitive IoT applications \cite{mao2017survey}. The reason is two-fold. First, the number of IoT devices is exponentially increasing and it is expected that it will be approximately 125 billion by 2030 \cite{abusafia2020reliability}. These devices generate a large volume of network traffic that burdens the backhaul network and significantly affects its performance by network congestion \cite{xia2022constrained, lai2022dynamic}. Second, cloud servers are typically placed at a remote distance from IoT devices. As a result, cloud computing introduces a considerable amount of delay in service provisioning, which degrades the overall QoS of delay-sensitive IoT applications \cite{liu2021qoe, xu2020distance}.}

\textcolor{black}{
Edge computing is a relatively new paradigm that offers an alternative computing solution for delay-sensitive and resource-intensive IoT applications. Edge computing extends cloud computing technologies to the edge of a network, closer to users and IoT devices \cite{lin2020survey}. It allows a resource-limited IoT device (a.k.a., edge devices) to fully or partially offload its data and computational tasks to nearby powerful edge servers or other edge devices \cite{abbas2017mobile}. It substantially improves the latency and energy efficiency of IoT applications. This also reduces the traffic congestion at the core network. Edge servers also work as data caches to store frequently accessed data by IoT devices to improve the QoS of the applications \cite{zhao2021fairness, zhou2023data}.}

\textcolor{black}{
IoT devices are typically connected to an edge infrastructure using wireless networks \cite{mao2017survey}. However, a good wireless network infrastructure may not always be available in some of the remotest areas, e.g., rural or mountain \cite{ji2020energy}. Moreover, a wireless network infrastructure can easily be affected by natural disasters such as earthquakes, floods, or storms. In military operations or emergency rescue missions, it may often be difficult to have a trustworthy wireless network infrastructure \cite{lewis2014tactical}. Unmanned Aerial Vehicle (UAV) technologies have opened a new opportunity where edge computing services are provisioned using UAVs in military operations, disaster response, or rural areas. This is also known as \textit{UAV-enabled edge computing} (UEC) \cite{mei2019joint}. UAVs offer a wide range of befits such as mobility, flexibility, and cost-efficiency, which make UEC a promising solution. A UAV typically works as an aerial edge server or a relay in a UEC environment \cite{li2020energy}. IoT devices offload their computational tasks fully or partially to a nearby UAV. A UAV either processes the tasks locally or sends them for remote execution to nearby edge/cloud servers.}     

\subsection{Resource Management in Edge Computing and UEC}
\label{subsec:resource_management}

Resource management in edge computing has been investigated intensively by many researchers in the last decades \cite{xia2023olmedc, liu2019survey}. Edge computing offers many unique advantages, such as distributed frameworks and low-latency services, however, it poses many unique challenges, including resource provisioning, computation offloading, and resource allocation.

In general, resource management refers to the use of a set of actions and methodologies to allocate resources, such as bandwidth, energy, CPU, GPU and many others, to devices or tasks for achieving specific objectives, i.e., maximizing utility, minimizing service latency, etc \cite{he2022pyramid, li2022edgewatch, liwang2022unifying, liwang2022overbook, xia2022formulatingcost}. In edge computing, the resource has its new characteristics: 1) limited - the resources on an edge server are usually constrained due to the limited physical size with smaller processors and a limited power budget \cite{chen2018label, li2020read, jin2022cost}; 2) dynamic - task requirements, device mobility, available resource amounts on edge servers are always changing over time \cite{xia2021online}; and 3) heterogeneous - both resources on edge servers and resources requirements of devices and tasks are heterogeneous \cite{pasteris2019service}.

Edge computing provides storage, computing and communication resources at the network edge to improve serviceability significantly. To ensure the utility of edge resources, appropriate resource scheduling strategies are required from two perspectives: infrastructure providers and services providers. Such resource management problems have been intensively investigated from the perspective of infrastructure providers with different optimization objectives, e.g., minimum transmission cost \cite{tran2018adaptive}, maximum system throughput \cite{deng2020wireless}, maximum data sharing efficiency \cite{luo2019software}, etc. From a service provider's perspective, the objectives of resource management can be minimizing resource rent cost \cite{xia2019graph}, maximizing users' QoS/QoE \cite{lai2019edge}, maximizing users' total data rate \cite{xia2022data}, maximizing its revenue \cite{samanta2019adaptive}, and many others. 

Resource management in UEC is a topical research area, attracting many researchers from both academia and industry. \cite{zhou2020mobile,liu2020unmanned}. {Table \ref{tab:paper_number} depicts the number of papers related to resource management in UEC published in the last 5 years. The details of the literature methodology are explained in Section \ref{subsec:literature_methodology}.} It clearly shows an increasing interest in the research community's inefficient resource management in UEC. Compared with traditional edge computing, UEC offers several advantages such as flexibility and autonomy with the fast deployment of UAVs for serving devices in rural areas or for critical missions \cite{feng2020joint, wu2018capacity, wu2018common}. However, UEC also poses many new challenges with its unique characteristics, such as multiple roles of UAVs, the mobility of servers, and the UAV trajectory design. Specifically, UAVs can play multiple roles when formulating resource management strategies in UEC, including servers, relays and users. The high mobility and limited resources of UAVs also complicate resource management in UEC. In addition, external factors, such as wind, rain and storm, can impact the performance of UAVs significantly. Considering those complicated scenarios in UEC, researchers from academia and industry have started to investigate to identify and develop suitable strategies for resource management in UEC. To guide researchers in this area, we have conducted this survey to provide an overview of resource management in UEC and point out the challenges that can direct future research.

\subsection{Related Surveys}


\begin{table} \renewcommand{\arraystretch}{1.0}
\footnotesize 
\caption{{Number of related papers}}
    \centering
    \begin{tabular}{|p{.8cm}| p {1cm}| p {1cm}| p{1cm} | p{1cm} | p{1cm} | p{1cm} |}
    \hline
    Year & 2017 & 2018 & 2019 & 2020 & 2021 & 2022 \\
    \hline
    \hline
    Number & 2 & 14 & 28 & 58 & 67 & 96 \\
    \hline
    \end{tabular} 
    \label{tab:paper_number}
\end{table}


    

Existing studies have focused on different aspects of resource management in UEC including resource provisioning and allocation, task or data offloading, and trajectory design. \textit{However, there is no existing survey or review that provides a comprehensive overview of resource management in UEC.} A few survey studies of UEC are mainly limited to the computational offloading aspect of resource management in UEC. A survey on offloading in UEC is presented in \cite{huda2022survey}. This study reports a comparative assessment of various offloading algorithms based on their performance and features. The survey provides different application scenarios and a case study of where UAVs could be leveraged to enable edge computing. Examples of such applications are next-generation wireless networking, surveillance of property, UAV-enabled target tracking, UAV-MEC in a pandemic, and reconnaissance in military operations. It provides a classification of offloading algorithms based on offloading policies such as binary offloading, partial offloading, and relay. 

\begin{table} \renewcommand{\arraystretch}{1.0}
\footnotesize 
\caption{A list of recent surveys in UAV-enabled Edge Computing}
    \centering
    \begin{tabular}{|p{.5cm}| p {4.2cm}| p {0.5cm}| p{7.4cm} |}
    \hline
    Paper & Title & Year & Key Contributions \\
    \hline
    \hline
    
    \cite{zhou2020mobile} & Mobile Edge Computing in Unmanned Aerial Vehicle Networks & 2019 & The paper presents three UEC architectures based on the role of UAVs in UEC. It provides a brief survey on computational offloading and resource allocation in UEC. \\
    \hline
    
    \cite{abrar2021energy} & Energy Efficient UAV-Enabled Mobile Edge Computing for IoT Devices: A Review & 2021 & The paper presents a brief survey on UEC networks with a focus on energy efficiency. It introduces basic terminologies and architectures used in UEC. It also presents various techniques and challenges related to computational offloading and a brief overview of resource management in UEC.  \\
    \hline
    
    \cite{yazid2021uav} & UAV-Enabled Mobile Edge-Computing for IoT Based on AI: A Comprehensive Review & 2021 & This paper presents a review of deep learning and machine learning techniques used in various applications of UEC. \\
    \hline

    \cite{huynh2022uav} & UAV-Enhanced Edge Intelligence: A Survey & 2022 & The paper introduces and presents various applications of UAV-enhanced edge intelligence systems. It also discusses some challenges and future research directions for UAV-enhanced Edge Intelligent systems. \\ 
    \hline

    \cite{huda2022survey} & Survey on computation offloading in UAV-Enabled mobile edge computing & 2022 & The paper presents a comprehensive survey of computational offloading research in various UEC systems. It compares existing algorithms qualitatively to assess their features and performances.  \\
    \hline          

    {\cite{song2022comprehensive}} & {A comprehensive survey on aerial mobile edge computing: Challenges, state-of-the-art, and future directions} & {2022} & {The paper provides an extensive survey of UAV optimization problems of UAV-enabled mobile edge computing with applications of Machine Learning techniques.}  \\
    \hline    

    {\cite{singh2023survey}} & {A survey of mobility-aware Multi-access Edge Computing: Challenges, use cases and future directions} & {2023} & {This paper comprehensively reviews the service migration, task offloading, resource allocation and content caching problems in EC. The paper briefly explores the role of UAVs as user equipment.} \\
    \hline

    \end{tabular}
    
    \label{tab:surveys}
    
\end{table}

A short survey is reported in \cite{abrar2021energy} which provides an overview of UEC networks and focuses on energy efficiency-related studies. It summarizes key computational offloading techniques concerning the ground mobile users who offload computational tasks to nearby UAVs. It also provides a summary of the approaches for the resource management of UAVs from the energy efficiency aspect. A survey on UAV-enhanced edge intelligence is presented in \cite{huynh2022uav} where edge intelligence is introduced as a key concept in fifth-generation networks. The survey presents various applications of UAV-enabled edge intelligence such as smart agriculture, disaster relief, and intelligent transport systems. The survey also discusses some key challenges to realising UAV-enabled edge intelligence such as energy efficiency, practical implementation, security, and privacy. A comprehensive review is carried out in \cite{yazid2021uav} which focuses on the role of Machine Learning, and Deep Learning based methods to enable UEC. The review presents an extensive study of different types of UAVs, their capabilities, and possible applications. Examples of the application areas are agriculture, industry 4.0, environment monitoring, health and emergency, smart cities and smart homes. The review also presents a classification of the relevant studies in UEC in terms of AI-based approaches. {There are also other surveys in UEC from different perspectives, such as security \cite{gupta2022secured}, UAVs as user equipment \cite{singh2023survey}, machine learning applications\cite{song2022comprehensive}, etc.} \textit{Different from the existing surveys, our prime focus is to provide a holistic view of resource management in UEC based on the existing relevant studies.} To the best of our knowledge, there is no existing review that provides a comprehensive overview of resource management for UEC. A list of the recent surveys is shown in Table \ref{tab:surveys}.

\subsection{{Literature Methodology}}
\label{subsec:literature_methodology}

{The selection of studies for our research was guided by the principles outlined in the Systematic Literature Review guidelines \cite{keele2007guidelines}. Here, we first identified and designed the keywords to retrieve papers from widely-used databases including ACM Digital Library, IEEE Xplore, Wiley, Scopus, SpringerLink, ScienceDirect and Hindawei: \textit{"'Unmanned Aerial Vehicle' OR UAV, Edge OR Fog, offloading OR workload OR scheduling OR share OR allocation OR resource OR task OR schedule OR sharing OR cache OR data OR storage OR provision OR position OR trajectory". }
}

{\textbf{Study selection}. We obtained 277 papers using the keywords above and defined inclusion/exclusion criteria based on \cite{keele2007guidelines} to filter out irrelevant or unnecessary studies. Based on these criteria, we removed 162 papers from the initial set of 277. After reading the full text and applying the criteria, we obtained 50 papers directly related to resource management in UEC. To further increase study coverage, we performed backward and forward snowballing on these 50 papers, identifying 7 more papers. We included a total of 57 studies for our survey. Acknowledging that our selection may not encompass all relevant literature in the field, we are confident that our selection encompassed the majority of essential research studies that reveal the techniques for managing resources in UAV-enabled edge computing.}

{\textbf{Data extraction and synthesis}. To ensure a thorough analysis, we conducted a pilot study of 10 papers to gain familiarity with the data to be extracted from the primary studies. Using initial codes, we iteratively merged them in several rounds to create themes. The analysis was carried out independently by two authors, with each author analyzing half of the selected papers and then reviewing the analysis output of the other author. Cases of disagreement were resolved through discussions among all the authors.}

\subsection{Contributions and Organization}

\begin{figure}
    \centering
    \includegraphics[width=.8\textwidth]{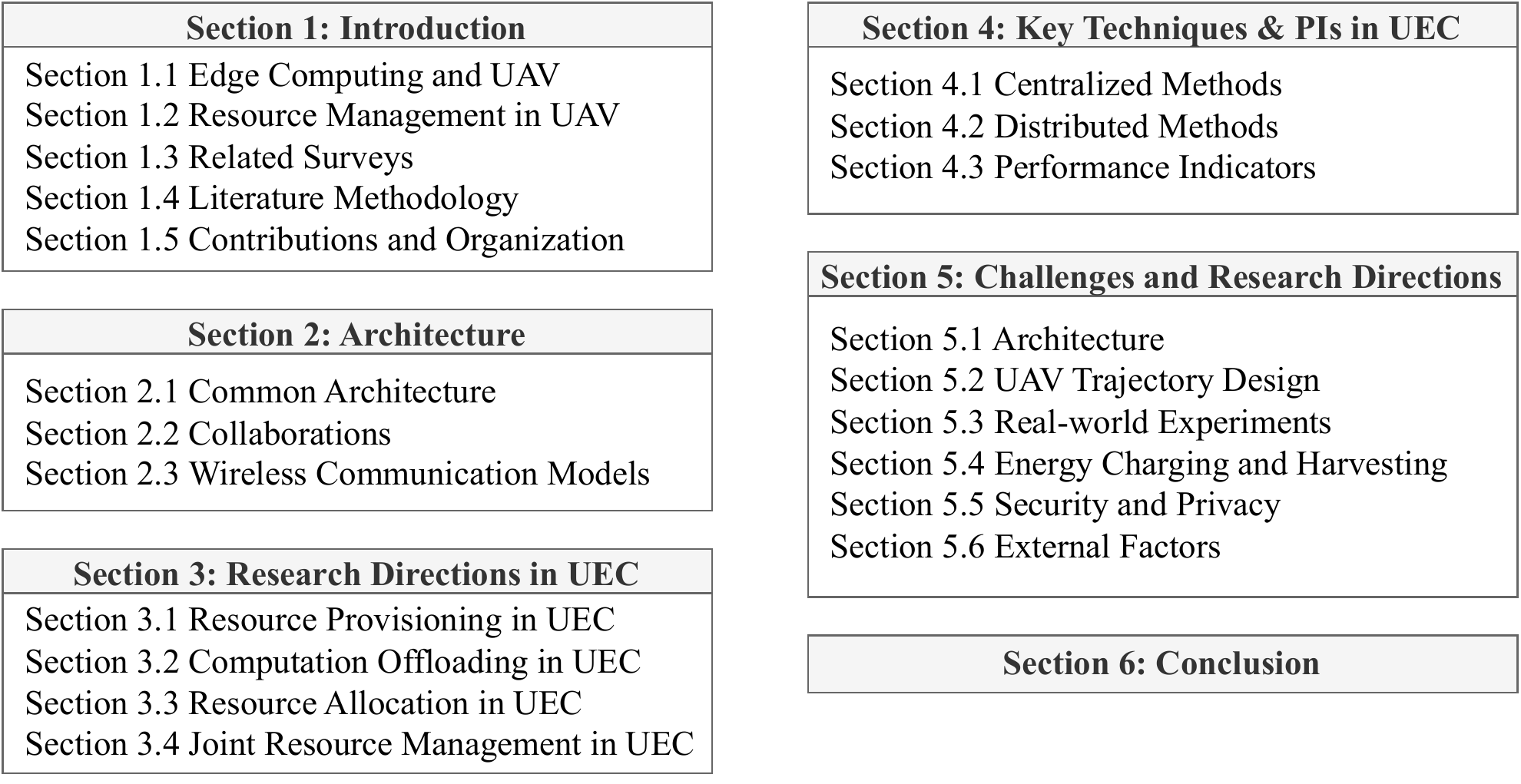}
    \caption{{Organization of the Survey}}
    \label{fig:paper_org}
\end{figure}

\textcolor{black}{
This survey provides a comprehensive review of the state-of-the-art research on resource management in UEC. The key contributions of this work are as follows. {\ding{182}} We introduce a three-layered UEC architecture in section 2, representing a conceptual architecture for managing resources in UEC. The architecture contains a Things layer, an Edge layer, and a Cloud layer. We then investigate six types of collaborations that take place in the proposed architecture. The considered collaborations are a) Things-UAV, b) UAV-Edge, c) Things-Edge, d) Things-UAV-Cloud, e) UAV-Edge-Cloud, and f) Thigns-UAV-Edge-Cloud. We also discuss the wireless communication models used in UEC. {\ding{183}} We discover the key research problems of resource management in the context of UEC. In Section 3, we categorize the research problems into the three following categories: a) computational tasks and data offloading, b) resource allocation, and c) resource provisioning. {\ding{184}} Section 4 identifies and comprehensively reviews the key techniques and performance indicators used for resource management in UEC. The key techniques are categorized into two categories: a) centralized methods and b) decentralized methods. We investigate how these methods are evaluated in the existing work. In addition, The key performance indicators such as energy consumption, latency, throughput, cost, utility, and resource utilization in the existing literature are discussed. {\ding{185}} We identify our key findings from this work in section 5 which points out the key research challenges and the future research directions in UEC for resource management.
}

Figure \ref{fig:paper_org} illustrates the organization of this survey to provide the readers with a brief overview of this paper.

\section{Architecture}

\textcolor{black}{
We present a conceptual architecture based on the existing UEC-related studies for resource management in this section. A typical UEC architecture consists of three layers including things, edge, and cloud. Existing studies investigate different types of scenarios for resource management in a UEC environment. As a result, different types of collaborations between things, UAV, edge, and cloud can be found in the existing studies. We highlight different types of collaboration between things, UAV, edge, and cloud in a typical UEC environment. In addition, we analyze and discuss the advantages and limitations of these collaborations under different contexts. Finally, we discuss the wireless communication models used in UEC systems.}

\subsection{Overview of the Conceptual Architecture}

\begin{figure}
  \centering
  \includegraphics[width=0.7\textwidth]{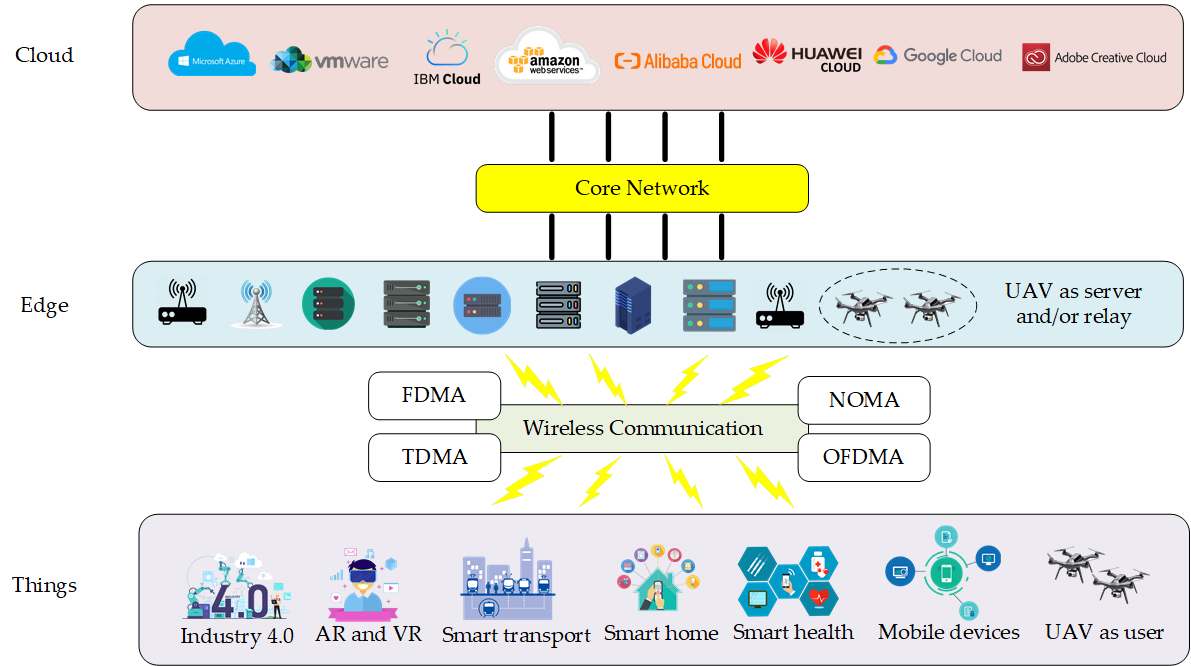}
  \caption{A Conceptual Architecture for Resource Management in UAV-enabled Edge Computing Environment}
  \label{fig:architecture}
\end{figure}

Figure \ref{fig:architecture} illustrates a typical three-layer UAV-enabled edge computing structure. The bottom layer, called things, consists of end-devices such as smart devices and autonomous vehicles, and different applications such as smart agriculture, smart home, smart health and Industry 4.0 \cite{liu2020unmanned}. End-devices are often called user devices (UD), mobile devices (MD) or user equipment (UE). These devices continuously collect data from the environment, which requires processing to support smart applications. These devices typically have limited resources and energy. Data collected by these devices are either processed locally or offloaded to the UAVs based on available resources and required QoS. The existing studies develop numerous offloading approaches to efficiently manage the resources in the things layer based on required QoS, i.e., latency, energy efficiency and throughput \cite{huda2022survey}. In a typical UEC environment, end-devices can be fixed in a particular location or change their position over time such as mobile users or autonomous vehicles. Based on the mobility of the devices, different types of offloading strategies are proposed in the existing literature.


The layer above the things layer is referred to as the edge layer, which comprises base stations, gateways, edge servers, cloudlets, and UAVs. UAVs in this layer can either offer edge services directly to the things layer or function as relays for offloaded tasks or data sent to edge servers or the cloud. The edge layer typically delivers a range of edge services, such as computing and storage services, to the things layer through wireless communication technologies like 4G and 5G. Effectively managing resources within the edge layer of UEC necessitates efficient resource allocation and provisioning of UAVs \cite{liu2020cooperative}. Due to the limited resources of UAVs, task offloading from UAVs to other UAVs, edge servers, or the cloud is often required to fulfil the QoS requirements of the things layer.

The cloud layer includes various public and private clouds such as Amazon, Microsoft, IBM, and Google, supporting the edge layer via the core network \cite{luo2021resource}. The cloud layer consists of large data centres, powerful computing capabilities, and high-speed communication resources. The edge layer often relies on the cloud layer to manage resource-intensive tasks and global information.

\subsection{Collaborations}

\begin{table}\renewcommand{\arraystretch}{1.0}
\footnotesize 
    \centering
    
    \caption{Collaborations and Roles of UAV in UEC}
    \begin{tabular}{|p{4cm}| p {5cm}| p {3cm}|}
    \hline
         Collaboration Types & Related Studies & Role of UAV \\
         
    \hline
    \hline
    Things - UAV &  \cite{feng2021hybrid}, \cite{yang2019energy}, \cite{ye2020offspeeding},  \cite{shimaday2020novel},   \cite{wang2021cooperative}, \cite{xu2020big}, \cite{wang2021computation}, \cite{wang2019joint}, \cite{ning2021dynamic}, \cite{sun2021joint}, \cite{hu2018joint}, \cite{khochare2021heuristic}, \cite{zhang2019computation},  \cite{zhao2021fairness}, \cite{liu2021joint},    \cite{luo2021optimization}, \cite{jeong2017mobile}, \cite{li2020energy}, \cite{ji2020energy},  \cite{ouyang2021trust}, \cite{gu2021uav},  \cite{bai2019energy}, \cite{qian2022joint}, \cite{wang2022bandwidth}, \cite{xu2022stochastic} & Edge Server     \\ 
    \hline 
    
    UAV - Edge & \cite{liwang2021let}, \cite{xu2021edge},  \cite{dai2020energy}, \cite{guo2021coded}, \cite{zhu2022auxiliary}  & User  \\

        \hline 
    
    Things - UAV - Edge & \cite{wang2021resource}, \cite{wu2020cell},  \cite{cheng2018uav}, \cite{he2021multi}, \cite{hu2020wireless}, \cite{apostolopoulos2021data}, \cite{zhang2020energy},  \cite{chen2020age}, \cite{liu2020joint}, \cite{yu2020joint}, \cite{liu2020cooperative}, \cite{dai2021towards}, \cite{hu2019task}, \cite{liu2019uav}, \cite{seid2021multi}, \cite{xiao2022resource}, \cite{zheng2022service}    & Edge Server and Relay  \\ 
    
    \hline
    
    Things - UAV - Cloud &  \cite{wu2021joint}, \cite{mao2020joint}, \cite{wan2019toward},  \cite{mei2019joint}, \cite{ti2018joint}, \cite{xie2022providing} & Edge Server and Relay \\ 
    
    \hline 
    
    Things - UAV - Edge - Cloud & \cite{asheralieva2019distributed}, \cite{zhang2020balancing}, \cite{liu2020online}, \cite{xu2022edgeworkflow} & Edge Server and Relay \\ 
    
    \hline 
    
    \end{tabular}

    \label{tab:collab}
\end{table}

Table \ref{tab:collab} demonstrates five types of collaborations between things, UAV, edge, and cloud that are considered in the existing literature. These collaborations are Things - UAV, UAV - Edge, Things - UAV - Edge, Things - UAV - Cloud, and Things - UAV - Edge - Cloud.

\textcolor{black}{
\subsubsection{Things - UAV} Resource management in the Things-UAV layer considers the interaction between the Things layer and the UAVs in the edge layer. Most studies consider that one or multiple UAVs are deployed to enable UEC in areas where the terrestrial network is unreachable or insufficient to maintain the required QoS. UAVs are typically considered flying edge servers in most existing studies where the Things-UAV collaboration is the key focus.
For example, a fairness-aware approach is proposed for resource management in \cite{zhao2021fairness}. The proposed approach considers that a UAV works as a flying edge server and provides computing and caching services to the ground nodes. Efficient resource management of UAVs is the key concern in this case to improve various QoS attributes. The proposed approach minimizes the energy consumption of a UAV by considering its trajectory and resource allocation. 
A joint resource scheduling approach is proposed where a UAV provides offloading services to a group of vehicles. The vehicles are considered capable of wireless power transmission. The goal is to increase the computational throughput with communication and computational resource constraints.
Depending on the number of available UAVs, different types of resource management strategies and optimizations are required. For instance, a unified framework is proposed in \cite{luo2021optimization} that improves the energy efficiency for ground users when multiple UAVs are deployed to enable a UEC system. The proposed framework jointly optimizes task scheduling, bit allocation, and UAV trajectory using a two-layer optimization strategy.
An energy consumption minimization approach is proposed in \cite{ji2020energy} where the energy consumption of UAVs and ground nodes is optimized by considering the UAV trajectory and its resource allocation.}

\subsubsection{UAV - Edge}

The architecture in Fig. \ref{fig:architecture} shows that UAVs can be either in the Things or Edge layer. Several studies consider UAVs as moving users with limited resources. Under the UAV-Edge collaboration, UAVs are normally required to offload tasks to the nearby edge servers to pursue their optimization goals, e.g., maintaining their desired QoS, maximizing the throughput, etc. To name a few, Liwang et al. propose a resource trading mechanism between a UAV (buyer) and an edge server (seller) \cite{liwang2021let}. Since the edge server has to serve users in its coverage, the available resources that can be provided to a UAV, are always limited and dynamic. The mechanism proposed in this paper considered both contracts for future resource trades and transmission power optimization. To tackle the resource trades between multiple edge servers and multiple UAVs, a Stackelberg game-based algorithm is introduced in \cite{xu2021edge}. In this study, edge servers provide the price of resources, and the UAVs make decisions to hire resources from edge servers to maximize utilities. In \cite{dai2020energy}, UAVs offload the computation tasks collected from nearby devices to edge servers, similar to \cite{guo2021coded}. 

\textcolor{black}{
\subsubsection{Things - UAV - Edge} UAVs are often deployed to extend edge computing services in remote areas for sparsely distributed ground nodes. However, UAVs are subject to limited resource capacity. Many existing studies leverage edge servers located at nearby base stations. This type of arrangement enables the Things - UAV - Edge collaboration. In this collaboration, UAV may work as an aerial edge server or a relay that forwards offloading tasks to nearby edge servers. For instance, a Things - UAV - Edge collaboration is proposed in \cite{yu2020joint} where a UAV is placed to enable edge services to ground users. The paper assumes that users cannot communicate directly with the nearby base stations or Wi-Fi access points. Therefore, the paper proposes a joint optimization approach that considers task offloading, resource allocation, and UAV trajectory to improve the service latency for all ground nodes and the energy efficiency of the UAV. In \cite{liu2020joint}, a cluster of UAVs serves a set of mobile users. This scenario considers that a UAV can offload tasks to nearby UAVs to maintain the ground users' expected QoS. A similar scenario is studied in \cite{liu2020cooperative}, where a cooperative scheme is proposed to efficiently carry out the computational offloading process in a UEC system. The goal of the proposed scheme is to maximize the long-term utility of UEC using an interference mitigation technique. A task and bandwidth allocation approach is proposed in \cite{hu2019task}, where a UAV acts as either a relay or an edge server. The proposed approach aims at maximizing the energy efficiency of the UAVs and ground nodes. It proposes a joint optimization technique that optimizes the task allocation, bandwidth allocation, and the UAV's trajectory.} 

\textcolor{black}{
\subsubsection{Things - UAV - Cloud} Several studies have considered a Things - UAV - Cloud collaboration where devices in the Things layers in the remote and mountainous areas without nearby cellular or base stations. Therefore, UAVs are deployed to enable edge services and satellite-based services by providing ubiquitous and seamless network coverage. For example, a space-air-ground integrated network is considered to provide computational offloading services for remote IoT applications to achieve the min-max service latency \cite{mao2020joint}. In this work, a set of UAVs are deployed to provision various IoT and mobile devices in remote areas. The LEO satellite-based cloud is envisioned that provide ubiquitous coverage. Computational tasks of the devices are partitioned to be executed by UAVs and LEO satellites. Similarly, an online data processing network with this three-layer collaboration is considered in \cite{wan2019toward}, where UAVs are deployed as moving edge servers that collect data from distributed sensors and conduct the initial processing. UAVs send these results to a centralized cloud for further processing. A similar scenario is studied in \cite{mei2019joint}, where the concept of virtualized network function is leveraged to help ground nodes in offloading tasks. A joint resource allocation and UAV trajectory optimization approach is proposed to minimize the overall energy consumption of a UAV. A hierarchical edge-cloud is considered where a set of users, a set of UAVs, and a macro base station provision UEC \cite{ti2018joint}. In this case, the study assumes that each ground node can offload its computation task to at most one cloud or cloudlet server.}

\subsubsection{Thing - UAV - Edge - Cloud}
Some studies have considered a Things - UAV - Edge - Cloud collaboration in which UAVs work as a relay and/or server to serve devices in the Things layers, collaborating with edge servers and remote cloud servers by borrowing resources from them. This is the most comprehensive architecture of UEC. In \cite{asheralieva2019distributed}, UAVs collect tasks from devices acting as a relay and then send those tasks to edge servers. Once edge servers receive those tasks, each edge server checks its capacities for task execution and offloads partial tasks to the cloud if it cannot handle all the tasks. Similarly, In \cite{liu2020online, zhang2020balancing}, UAVs collect data or tasks from different areas and then send them to edge servers for execution. If there are not enough resources, part of the data or tasks will be transferred to a remote cloud. In this scenario, the capacities of UAVs and edge servers can be fully utilized and the resources in cloud servers could be borrowed to complete all the tasks from devices with consideration of limited resources on UAVs and edge servers. The evidence that many existing studies do not fully collaborate UAVs, edge servers and cloud servers for serving devices, indicates that more efforts are required when researchers investigate resource management problems for formulating a comprehensive architecture.

\subsection{Wireless Communication Models}
\label{subsec:common_models}

In the edge computing environment, especially UEC, wireless communication plays a significant role in connecting devices, UAVs and edge servers. Multiple devices can access UAVs and edge servers for data and resources via wireless communication concurrently. The wireless interference incurred may significantly impact their achievable data rates and lower their quality of experience \cite{chen2016efficient, xia2022data}. Thus, the interference must be considered during resource management in UEC.
There are two major schemes of wireless communication between devices and servers, i.e., Orthogonal Multiple Access (OMA) and Non-Orthogonal Multiple Access (NOMA). 

Under the OMA scheme, the resources, i.e., frequency and time, can be divided into sub-resources to achieve the minimum interference between servers and devices. The most popular OMA schemes in existing studies are Time Division Multiple Access (TDMA), Frequency Division Multiple Access (FDMA), and Orthogonal Frequency-Division Multiple Access (OFDMA). Connectivity, i.e., multi-device access to the radio resources in cellular systems including the Global System for Mobile Communication, the Long Term Evolution and Long Term Evolution-Advanced is ensured with OMA-based multiple access. 
As a new multiple access scheme implemented in 5G, NOMA allows non-zero cross-correlation signals. In this way, NOMA provides connectivity for massive devices and enhances the spectral efficiency significantly, compared to conventional OMA schemes \cite{ding2017survey}. In the edge computing environment, each edge server deployed with a base station is equipped with multiple channels, and each channel is available for multiple device accommodation at the same time while guaranteeing data rates by well-designed transmit power allocation \cite{fu2019joint, liu2017highly}. The successive interference cancellation is implemented in the NOMA scheme to increase data rates of devices with poor channel conditions by treating the signals of devices with better channel conditions in the same channel as noise \cite{fu2019joint}. 

Apart from the OMA or NOMA schemes applied in UEC, the line-of-sight (LoS) also needs to be considered when formulating resource management strategies. This is because UAVs hover at high altitudes in the air to serve devices on the ground, and can use the LoS links for data and task communication \cite{luo2021optimization, tun2020energy, deng2021energy, zhou2018uav, chen2020joint}. Compared with traditional download and upload links, LoS links are dominant over other communications due to their small-scale shadowing and fading \cite{abrar2021energy}. 
According to the communication protocol or scheme selected in UEC scenarios, we can calculate the data rates of upload and download links. For computation, energy-consumption and latency models, we omit them here because those models have been clearly defined in related surveys in edge computing such as \cite{luo2021resource, mao2017survey}.

\section{Research Directions in UEC}



\begin{figure}
    \centering
    \includegraphics[width=\textwidth]{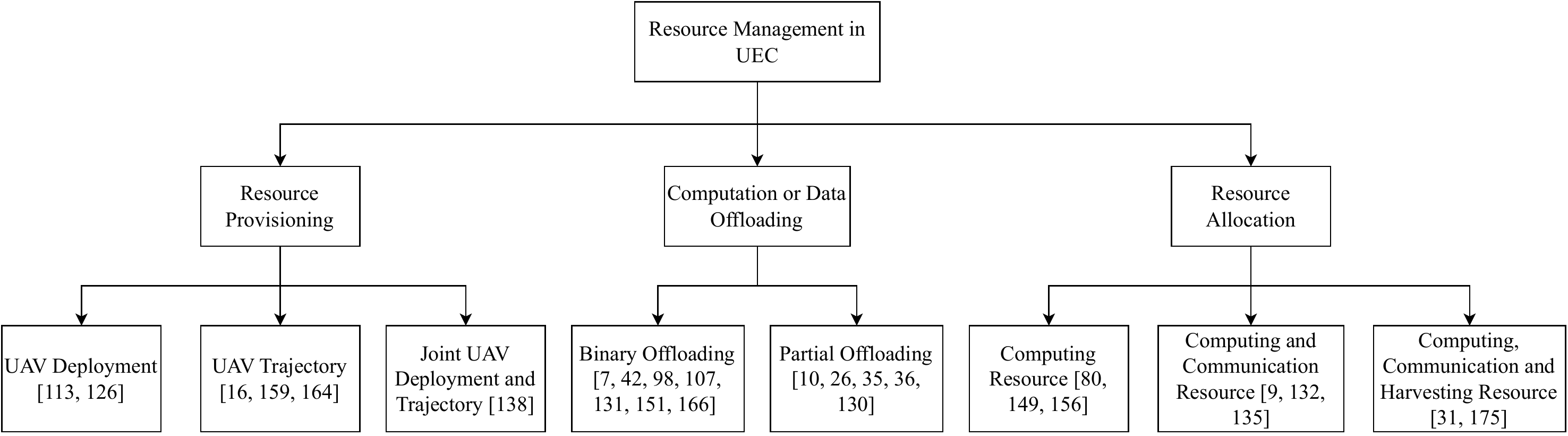}
    \caption{{A taxonomy of key research directions in UEC for resource management}}
    \label{fig:research_problem}
\end{figure}

Efficient resource management is the key to unlocking the full potential of UEC. Resource management is therefore extensively studied in the existing literature. We have identified three main aspects of resource management in UEC as shown in Fig. \ref{fig:research_problem}. Firstly, UEC needs to deploy UAVs in a geographic area to provide resources for potential services and/or tasks. Considering the mobility of UAVs, the trajectory design should also be designed. Those are referred to as \textit{resource provisioning}. After that, the second aspect is \textit{computational offloading} required by various entities in the Things layer (Figure \ref{fig:architecture}). Efficient offloading strategies require intelligently deciding where to offload and how much data or tasks to offload. The last aspect is to allocate resources provisioned for offloaded data or tasks. \textit{Resource allocation} requires efficient allocation of communication, storage, and computational resources in Things, Edge, and Cloud layers.

\subsection{Resource Provisioning in UEC}
\label{subsec:provsioning}


Compared with servers in cloud computing, the servers, i.e., UAVs and edge servers, are physically much smaller and easier to deploy and re-deployed in UEC. In this way, resource provisioning, as the foundation research problem of resource management in UEC, should be adaptive for dynamic user requests over time. There can be under-provisioning or over-provisioning issues as a result of unsuitable resource provisioning strategies. In under-provisioning, i.e., less available resources on UAVs and edge servers, it is hard to fulfil users' Quality-of-Service (QoS) requirements, suffering huge delays and incomplete services. However, over-provisioning would result in resource waste without any benefit. Thus, it is vital to formulate suitable resource provisioning strategies to dynamically adjust the requirements, such as low system cost, high QoS and/or QoE.

Different from traditional edge computing, UAV plays a significant but flexible role in UEC. Thus, when investigating resource provisioning problems in UEC, UAV behaviours must be considered. Since the edge server deployment is not unique in UEC, we will elaborate on the UAV deployment, UAV trajectory and joint UAV deployment and trajectory in this section.

\subsubsection{UAV Deployment}

In a number of the existing studies, UAVs are equipped with small-scale edge servers with low-capacity computing and storage resources, called UAV-mounted edge servers. Equipped with such a UAV-mounted edge server, a UAV can hover at a specific location to provide services to devices within its wireless communication coverage. Except for the coverage, the hover location of a UAV and the external factors at this location also impact the distance from devices, the data rate received by those devices and the power consumption of this UAV. Thus, it is important to deploy UAVs in suitable hover locations for maximizing the number of served tasks, maximizing the overall data rates received by devices, minimizing power consumption and other objectives. In \cite{wang2019joint}, the authors investigate the UAV deployment problem via dispatching UAVs to a suitable hover position for serving the maximum number of tasks from devices on the ground. Similarly, Sun et al. \cite{sun2020optimizing} study the multiple UAV deployment problem to minimize the overall task execution time. However, the UAV trajectory, consuming necessary energy and impacting the waiting time of tasks, is not considered in the above studies.

\subsubsection{UAV Trajectory}
As the UAV deployment decides the hover locations of UAVs, another key to the success of providing long-term and stable services is how to design the trajectory of UAVs to properly utilize both resource capacities on UAVs and the energy of UAVs in a dynamic manner. This is also a unique characteristic to be considered when managing resources in UEC, fundamentally different from transitional edge computing. In \cite{zhang2020balancing}, Zhang et al. study the UAV trajectory in UEC for minimizing energy consumption while maximizing benefits when scheduling the path of UAVs to collect data for processing. Similarly, the authors of \cite{ye2020offspeeding} design the UAV trajectory from the view of a multi-hop routing schedule. In this study, UAVs visit multiple positions for task collection and then visit the base station for transferring tasks to edge servers. If those tasks cannot be completed by UAVs and edge servers, the remaining tasks would be delivered from edge servers to the cloud data centre via the Internet. However, in those studies, only the optimization of flight is investigated without consideration of the number of dispatched UAVs and the selection of hover positions for providing better services with less energy consumption.

\subsubsection{Joint UAV Deployment and Trajectory}
A more comprehensive way to investigate resource provisioning in UEC is to consider both UAV deployment and trajectory together since UAVs not only deliver data or tasks but also serve devices. From the perspective of real-world implementations, it is also important to consider UAV deployment and trajectory at the same time. In \cite{wu2021joint}, the demand for vehicles is simulated based on the real-time traffic conditions predicted by the cloud server. According to both the number of UAVs to be dispatched and traffic status, the authors design a deep reinforcement learning approach to decide the hover positions and design the flight path of UAVs to those positions. However, the scheduling of the number of dispatched UAVs is not considered in this paper. To reasonably utilize the resources of UAVs while minimizing overall costs such as energy consumption and UAV rent cost, researchers need to investigate the joint UAV deployment and trajectory problem more comprehensively, considering the number of involved UAVs, the hover position and corresponding time for serving devices and the UAV trajectory that visits required positions with high system throughput and low service latency. 

\begin{table*}
\renewcommand{\arraystretch}{1.0}
\footnotesize 
\caption{A Comparison Resource Provisioning Strategies in UEC}
\centering
\begin{tabular}{|c|p{2cm}|p{2.8cm}|p{2.5cm}|p{4cm}|}

    \hline
    Paper & Provisioning Type & Objectives & Solutions & {Benefit} \\
    \hline
    \hline
 

    \cite{wang2019online} & UAV deployment & Maximize the number of served tasks & Iterative algorithm & {Providing efficient solutions considering coverage constraint and task execution deadline in the dynamic UEC environment} \\
    
    \hline
    
    \cite{sun2020optimizing} & UAV deployment & Minimize task completion time & Successive convex approximation &  {Providing optimal solution considering limited computing power of UAVs} \\
    
    \hline
    
    \cite{zhang2020balancing} & UAV trajectory & Minimum energy consumption and maximum reputation gain & Game-theoretical approach &  {Ensuring the quality of user experience by guaranteeing data rate while considering the moving speed of the UAV} \\
 
    \hline
    
    \cite{ye2020offspeeding} & UAV trajectory & Minimize UAV's energy consumption & Heuristic algorithm & {Offering efficient solutions considering communication conditions in the dynamic UEC environment}  \\
    
    \hline
    
    \cite{cheng2018uav} & UAV trajectory & Maximize overall data rate & Iterative algorithm & {Offering efficient solutions considering limited energy of UAVs in the dynamic UEC environment} \\
    
    \hline
    
    \cite{wu2021joint} & UAV deployment and trajectory & Minimize energy consumption and maximize system reward & Genetic algorithm and deep reinforcement learning & {Proposing a combination approach to solve both UAV deployment and trajectory problems}  \\
    
    \hline

\end{tabular}

\label{tab:resource_prov} 
\end{table*}

\subsection{Computation Offloading in UEC}

Computational offloading is the prime motivation for enabling UEC and one of the most important research problems. It is therefore the most studied topic in resource management in UEC. In this subsection, we summarize the key research issues and the proposed techniques as solutions in the existing literature. 
To provide high performance with low latency in UEC, tasks or data should be offloaded to proper servers, i.e., edge servers and UAVs. In this case, there are two major types of offloading modes adopted in the existing studies, such as binary offloading and partial offloading. 

\subsubsection{Binary Offloading} In the UEC environment, binary offloading means that the tasks or data have to be locally executed or offloaded to UAVs, edge servers or the remote cloud as a whole. Since the decision can only be true or false, it is binary. This mode has been adopted in several studies,  \cite{apostolopoulos2021data, shimaday2020novel, xu2020big, wang2021cooperative, ouyang2021trust}, as it is easy to implement and suitable for relatively simple tasks. Apostolopoulos et al. \cite{apostolopoulos2021data} study a data offload problem in UEC where UAVs are additional servers to associate edge servers on the ground. Once a UAV collects data from a user, the UAV needs to determine whether to execute it locally or send it to edge servers. Aiming to maximize all users' satisfaction utility, the authors propose a game-theoretical algorithm. A similar problem has been studied in \cite{xu2020big}. In this work, UAVs are relays for creating communication between devices and edge servers when the nearby base stations have been destroyed in a disaster. {However, the situation studied in this study is ungrounded.} In \cite{shimaday2020novel}, the authors propose a workload balance model for UAVs while ensuring low service latency. In this model, the tasks can be either executed locally or transferred to another UAV via wireless communication. Another binary offloading study is proposed in \cite{zhang2020energy}. This study tackles the scenario that an edge server has been damaged in the wild while some devices need to be served. The authors propose a game-theoretical approach to offload the tasks from each un-served user to a newly involved UAV or to a nearby edge server that directly covers this user. {Unfortunately, this study failed to solve the complicated UEC scenarios with multiple UAVs and edge servers.} In \cite{ouyang2021trust}, the authors also deal with a binary offloading problem in UEC. A trust-based task offloading scheme is proposed to maximize the energy efficiency and reliability of the offloading process. A cluster-based task offloading is proposed where tasks only need to route to any IoT devices within a cluster. {Similar to \cite{zhang2020energy}, this work can only deal with a single UAV. In addition, for all the above studies, the binary offloading mode may waste idle resources and become the performance bottleneck of the overall system performance.}

\subsubsection{Partial Offloading} Compared with binary offloading, partial offloading allows a task to be partitioned into multiple sub-tasks. Part of those sub-tasks can be partially executed on the device, and others can be offloaded at different UAVs and edge servers. In this mode, the offloading is more complicated, since it requires task partition in advance. However, this mode is more effective than binary offloading, since it can utilize segmented resources on devices, UAVs and edge servers. In \cite{guo2021coded}, the authors use the coded distributed computing (CDC) scheme to divide the original task received by a UAV into a set of sub-tasks, and then distribute those sub-tasks to multiple edge servers for execution. In this way, task processes in UEC can be distributed and parallel. {However, this study ignores the edge server network and only considers the single edge server scenario.} The authors of \cite{he2021multi} also consider the partial offloading mode in UEC. In this study, the tasks are treated as a continuous data stream and once a UAV receives such steam, it directly executes part of the tasks on its computing resources. If there exist tasks waiting for execution, the UAV will offload them to other UAVs. A partial offloading approach is investigated in \cite{bai2019energy} where a UAV is considered as a user and offloads computational tasks to the nearby edge server. The study focuses on the three following research questions: a) What is the volume of the computational tasks to be offloaded? 2) what is the suitable duration of offloading, and c) How much power should be assigned to the offloaded signal? This study investigates secure offloading techniques focused on physical-layer security for both active and passive eavesdroppers and transforms them into convex problems. {Unfortunately, this study only focuses on a single UAV again, limiting its feasibility and implementation.} A similar study has been carried out in \cite{gu2021uav} where secure partial offloading is studied in a linear energy harvesting-enabled UEC environment. The study considers secrecy as an offloading constraint along with latency constraints to minimize energy consumption and maximize the energy storage of the UAV. The study focuses on data-partitioned-oriented applications where data are considered to be partitioned arbitrarily for parallel processing. {The improvement of this study could be considering non-linear energy harvesting scenarios.}  



\begin{table*}
\renewcommand{\arraystretch}{1.0}
\caption{A Comparison of Computational Tasks or Data Offloading in UEC}
\centering
\footnotesize 
\label{tab:offloadin}
\begin{tabular}{|c|p{3cm}|p{3.2cm}|p{4cm}|p{1.3cm}|}
    \hline
    Paper & Objective & Method & Constraints & Mode \\
    \hline
    \hline
 
     \cite{apostolopoulos2021data} & Maximize satisfaction utility & Game-theoretic approach & Users' latency and energy requirements & Binary \\
    
    \hline

    \cite{shimaday2020novel} & Minimize service delay & Integer programming solver &  Not mentioned &  Binary \\ 
    
    \hline 
    
    \cite{xu2020big} & Minimize service latency, and energy consumption of UAVs & Greedy approach & Not mentioned &  Binary  \\ 
    
    \hline 
    
    \cite{he2021multi} & Maximize network computing rate & Iterative algorithm & Bandwidth and UAV battery constraints & Binary \\ 
    
    \hline 
    
    \cite{wang2021cooperative} & Minimize overall task processing delay
 & K-means and convex toolkit & computation capacities and UAV battery constraint & Binary   \\ 
    
    \hline 
    
    \cite{zhang2020energy} & Minimize overall cost, consisting of latency and energy consumption
 & Game-theoretic approach & Not mentioned & Binary  \\ 
    
    \hline 
    
    \cite{guo2021coded} & Minimize overall cost, consisting of latency and energy consumption & Mixed Integer programming solver &  Fly speed & Partial \\ 
    
    \hline 
    
    \cite{wang2021computation} & Minimize maximum processing delay & Reinforcement Learning & UAV battery capacity and task deadlines &  Partial  \\ 
    
    \hline
    
    \cite{dai2020energy} & Maximize UAV offloading efficiency & Game-theoretic approach & Computing capacities of UAV and edge servers & Partial \\
    
    \hline 
    
    \cite{ouyang2021trust} & Minimize energy consumption of UDs & Cluster-based offloading, ant colony optimization & Trust, trajectory, energy, and flight duration & Binary \\
        
        \hline 
    
    \cite{gu2021uav} & Minimize energy consumption of UAV  & Convex optimization & Energy consumption, latency, and security  & Partial \\
    
    \hline 
    
    \cite{bai2019energy} & Minimize energy consumption of UAV & Convex optimization & Energy consumption, latency, and security & Partial \\ 
    
    \hline

\end{tabular}
\end{table*}

\normalsize

\subsection{Resource Allocation in UEC}

As a significant research issue of resource management in UEC, it is important to devise and thoroughly evaluate different resource allocation strategies effectively and efficiently for executing offloaded tasks. In general, computing and communication resources on devices, UAVs and edge servers are the main ones to be managed in UEC. Typical computing resources include CPUs, GPUs, memory, storage, etc., whilst the main communication resources refer to bandwidth, power, link and spectrum. Considering the power limit on devices and UAVs, power for harvesting has become a new research trend in resource management, different from power allocation in wireless communication. In this section, we discuss the resource allocation studies in computing, communication, harvesting and joint resource allocations.


\begin{table*}
\renewcommand{\arraystretch}{1.0}
\footnotesize 
\caption{A Comparison of Resource Allocation Approaches in UEC}
\centering

\begin{tabular}{|c|p{2.2cm}|p{1.3cm}|p{2cm}|p{1.5cm}|p{4cm}|}

    \hline
    
    Paper & Objective & Resource Direction & Resource Type & Method & {Benefit} \\
    \hline
    \hline
    
    \cite{liwang2021let} & Maximize both seller's and buyer's expected utilities & Edge server $\rightarrow$ UAV  & Computing resource & Bilateral negotiation algorithm & {Considering computation and storage resource constraints on edge server in the dynamic UEC environment} \\ 
    
    \hline 
    
    \cite{xu2021edge} & Maximize edge server's profit and maximize UAV's revenue & Edge server $\rightarrow$ UAV  & Computing resource & Stackelberg game and blockchain & {Solving problems in an online and distributed manner while using blockchain to enhance security}  \\ 
    
    \hline 
    
    \cite{yang2019energy} & Minimize power consumption of devices and UAVs & UAV $\rightarrow$ Things & Computing resource & Iterative algorithm & {Thouroughly considering coverage, computation capacity, communication power and computation power of the UAV and edge server} \\ 
    
    \hline
    
    \cite{wang2021resource} & Minimize overall system cost & UAV $\rightarrow$ Things & Communication and computing resources & Multi-agent reinforcement learning & {Considering bandwidth allocation with task failure penalty} \\ 
    
    \hline 
    
    \cite{wu2020cell} & maximize average spectral efficiency and overall network throughput & Edge server and UAV $\rightarrow$ Things & Communication and computing resources &  Threshold-based greed scheme & {Monitoring and managing UAV moving speed with coverage limitations in the dynamic UEC environment}  \\ 
    
    \hline 
    
    \cite{asheralieva2019distributed} & Minimize overall cost with low-latency  & Edge server and UAV $\rightarrow$ Things  & Communication and computing resources & Blockchain, Stackelberg game, and deep Q-learning & {Using game theory and deep learning together to ensure the performance of the proposed solution in the dynamic UEC environment while implementing blockchain to enhance trust and security} \\ 
    
    \hline
    
    \cite{zhou2018computation} & Maximize sum computation rate & UAV $\rightarrow$ Things & Computing, communication and harvesting resources & Iterative algorithm & {Utilizing energy harvesting technique to enhance the sustainable usage of UAV and end-devices with consideration of UAV fly speed limit} \\
    
    \hline 

    \cite{feng2021hybrid} & Maximize sum computation rate & UAV $\rightarrow$ Things  & Computing, communication and harvesting resources & Heuristic algorithm and convex toolkit & {Ensuring low-latency task completions while utilizing energy harvesting technique} \\ 
    
    \hline 
\end{tabular}
\label{tab:resource_allocation}
\end{table*}

\subsubsection{Computing Resource Allocation}

As mentioned above, resource allocation aims to optimize the computation offloading process for various goals, including QoS, service latency, energy consumption system utility, etc. Thus, computing resource allocation is key to the success of achieving effective computation offloading in UEC. Without proper computing resource allocation, the number of un-served or delayed tasks would increase dramatically, similar to under-provisioning discussed in \ref{subsec:provsioning}. Thus, many researchers have started to investigate how to effectively utilize computing resources provided by UAVs and edge servers in UEC. Liwang et al. \cite{liwang2021let} investigate the resource trading problem between a UAV and its nearby edge server. They design a beneficial and risk-tolerable forward contract for negotiation between the UAV and edge server before the real resource transaction happens. In \cite{xu2021edge}, the authors also study the computing resource allocation from edge servers to UAVs in UEC. However, this resource allocation scenario is more complicated, since there are multiple UAVs obtaining computing resources from nearby edge servers in the coverage of UAVs. Rather than allocating computing resources from edge servers to UAVs, the authors of \cite{yang2019energy} propose an energy-efficient resource allocation approach to allocate computing resources of UAVs to devices on the ground.
\textcolor{black}{
The main objective of this work is to improve the energy efficiency of multiple UAVs in a UEC system. It introduces a low-complexity algorithm that jointly optimizes power control, user association, resource allocation, and location planning.} 



\subsubsection{Computing and Communication Resource Allocation} 

As the "last mile" of delivering servers to devices in UEC, the impacts of wireless communication must be considered when investigating resource management problems in UEC. As mentioned in Section \ref{subsec:common_models}, multiple access schemes are enabled to power the 5G wireless network in UEC. In this way, the incurred interference can impact the QoS of users significantly by reducing received data rates, which widely exist in UEC. Thus, the communication resources, including bandwidth, power, link and spectrum, also need to be considered together with computing and storage resources to improve the service quality and performance. 
In \cite{wang2021resource}, bandwidth is the main communication resource that is considered when allocating computing resources from UAVs to devices in the Things layer. In this work, the latency requirements of tasks must be fulfilled to ensure the safety of devices, e.g., vehicles. To achieve this, the authors consider the computing resources, channel efficiency and bandwidth usage together as the optimization goal of this study. 
The authors of \cite{wu2020cell} consider the changes in communication statuses, e.g., channel gains, available power and bandwidth, between the UAV and devices on the ground. To provide high-quality offloading services to devices covered by the UAV flight, a theoretical framework is proposed in this paper to maximize the average spectral efficiency and overall network throughput. 
Asheralieva et al. \cite{asheralieva2019distributed} investigate the resource management and corresponding pricing problem in UEC. Because normally there are multiple infrastructure providers in the UEC environment, this work studies how to enable various providers to work together with incomplete information to suitably allocate available resources to minimize the overall cost with low service latency. 
 
\subsubsection{Computing, Communication and Harvesting Resource Allocation}
The energy harvesting techniques, e.g., wireless power transfer, have been treated as a potential solution for IoT devices to improve the battery performance \cite{tang2017energy, tang2018energy}. However, the heavy propagation loss can significantly degrade the harvested power level. To improve the efficiency of energy transfer, the UAV-enabled wireless power transfer architecture is proposed and widely adopted in recent studies \cite{zhou2018computation, feng2021hybrid, wang2017resource, xu2018uav}. Under this architecture, UAVs perform as a transmitter for powering devices nearby. Due to the short-distance line-of-sight (LoS) energy transmission links between UAVs and devices, the harvested power level is significantly improved. With the advantages of the integration of energy harvesting and UEC, the device battery life, service time and available energy for computation tasks can be largely extended. In \cite{zhou2018computation}, a multi-stage approach is proposed to maximize the computation rate in a UEC scenario by properly allocating computing resources, with consideration of energy harvesting and UAV flight speed constraints. Specifically, this study considers both binary and partial task execution modes with task completion deadlines. Feng et al. \cite{feng2021hybrid} study this joint resource allocation problem in UEC under the non-orthogonal multiple access (NOMA) scheme. Similar to \cite{zhou2018computation}, energy is also transferred from UAVs to devices by the joint technique of wireless power transfer and NOMA. To ensure the performance in both binary and partial task offloading, a deep deterministic policy gradient framework is proposed for allocating computing, communication and harvesting resources. 

\subsection{Joint Resource Management in UEC}

\begin{figure}
    \centering
    \includegraphics[width=.85\textwidth]{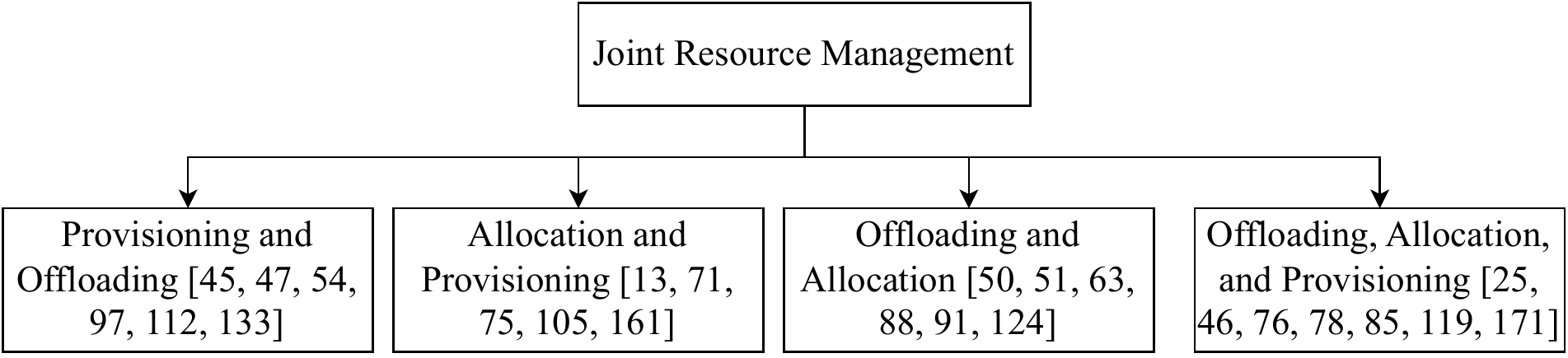}
    \caption{{A taxonomy of joint resource management}}
    \label{fig:join_resource_tax}
\end{figure}

\textcolor{black}{Many existing studies focus on resource management more than one aspect of resource management. We classify these studies into four categories based on their research focus as shown in Fig. \ref{fig:join_resource_tax}. The four categories are a) Provisioning and Offloading, b) Offloading and Allocation, c) Allocation and Provisioning, and d) Offloading, allocation, and provisioning. We provide an overview of the existing studies in these four categories in this subsection.}

\subsubsection{Joint Provisioning and Offloading}

When tackling the resource management in UEC with the special characteristics of UAVs, a number of the existing studies start to investigate both computation offloading and resource provisioning at the same time \cite{wang2019joint, ning2021dynamic, sun2021joint, hu2020wireless, hu2018joint, khochare2021heuristic}. In \cite{wang2019joint}, the authors consider the joint UAV deployment and task scheduling problem for large-scale users in UEC. The Things-UAV collaboration is adopted in this work to provide services to users by UAVs. This work aims to minimize the system energy consumption including computational energy consumption and hover energy consumption whilst ensuring all tasks can be completed. A similar study is carried out in \cite{ning2021dynamic} where UAVs act as servers to process tasks offloaded from devices on the ground. In this work, the optimization objective is to minimize the overall system cost, consisting of the energy consumed on UAVs and the latency of task completion. To solve this problem, a game-theoretical approach is designed to find the solution effectively and efficiently. {However, the coverage constraint, energy limit and UAV movement are not considered, weakening the novelty of this study.} Different from the above studies which consider UAV deployment and computation offloading, the authors of \cite{sun2021joint} study the joint UAV trajectory and computation offloading problem, with similar aims of \cite{ning2021dynamic}. Rather than the cost, the authors of \cite{hu2020wireless} focus on maximizing task completion and the authors of \cite{hu2018joint} focus on minimizing the delay over time when solving the joint UAV trajectory and computation offloading problem. {Unfortunately, both \cite{ning2021dynamic} and \cite{hu2018joint} only consider the single UAV scenario.} Khochare et al. \cite{khochare2021heuristic} investigate a co-scheduling problem of analytic tasks and UAV routing to maximize the overall utilities of UAVs and achieve the analysis completion before the deadline. {However, there is still a research gap that the UAV deployment, UAV trajectory and computation offloading at the same time have rarely been studied.}

\begin{table}[]\renewcommand{\arraystretch}{1.0}
\footnotesize 
    \caption{A comparison of Joint Resource Management Studies in UEC}
    \centering
    \begin{tabular}{|p{0.7cm}|p{8cm}|p{4.5cm}|}
     \hline
     Papers & Objectives & Research Focus \\
     \hline
    \hline
    
    \cite{wang2019joint} & Minimize system energy consumption while ensuring all tasks can be completed &  \multirow{6}{*}{Provisioning and Offloading} \\
    
    \cline{1-2}
    
    \cite{ning2021dynamic} & Minimize system computation cost, including latency and energy & \\
    
    \cline{1-2}
    
    \cite{sun2021joint} & Minimize UAV's energy consumption &  \\
    
    \cline{1-2}
    
    \cite{hu2020wireless} & Maximize weighted sum complete task-input bits &   \\
    
    \cline{1-2}
    
    \cite{hu2018joint} & Minimize the sum of maximum delay over time &  \\
    
    \cline{1-2}
    
    \cite{khochare2021heuristic} & Maximize utility & \\
    
    \hline
    

    \cite{chen2020age} & Maximize computation performance of UD & \multirow{5}{*}{Allocation and Offloading} \\
    
     \cline{1-2}

    
    
    
    \cite{yu2020joint} & Minimize the service delay of all UDs and energy consumption of UAVs &  \\ 
     \cline{1-2}
    
    \cite{seid2021multi} & Minimize computation cost  &  \\ 
    \cline{1-2}
    
    \cite{liu2020cooperative} & Maximize computation capacity &  \\
     \cline{1-2}
    
    \cite{liu2020joint} & Maximize Energy Efficiency &  \\
    
    \hline
    
    \cite{jeong2017mobile} & Minimize energy consumption of UDs & \multirow{6}{*}{Allocation and Provisioning} \\
     \cline{1-2}
    
    \cite{li2020energy} & Minimizing energy consumption of UAVs &  \\
     \cline{1-2}
    
    \cite{mao2020joint} & Minimize the maximum computation delay of UDs &  \\ 
     \cline{1-2}
    
    \cite{wan2019toward} & Enhance UAV coverage &   \\ 
    \cline{1-2}
    
    \cite{ji2020energy} & Minimize weighted-sum energy consumption UDs and UAVs &  \\ 
     \cline{1-2}
    
    \cite{mei2019joint} & Maximize the energy efficiency of an UAV &  \\ 
    
    \hline

    
    \cite{zhao2021fairness} & Minimize energy consumption of UAVs & \multirow{7}{*}{Offloading, Allocation and Provisioning} \\
    \cline{1-2}
    
    \cite{dai2021towards} & Minimize energy consumption of All UDs &  \\
    \cline{1-2}
    
    \cite{hu2019task} & Minimize energy consumption of UDs and UAVs &  \\
     \cline{1-2}
    
    \cite{liu2019uav} & Minimize energy consumption of UAVs &  \\ 
     \cline{1-2}
    
    \cite{ti2018joint} & Minimize energy consumption of UDs &  \\ 
     \cline{1-2}
    
    \cite{liu2021joint} & Maximize system-wide computation capacity & \\ 
     \cline{1-2}
    
    \cite{luo2021optimization} & Minimize user energy consumption &  \\ 
    \hline 
   
    \end{tabular}

    \label{tab:joint_resource}
\end{table}

\subsubsection{\textcolor{black}{Joint Offloading and Allocation}}

Several studies consider both offloading and allocation of resources to improve various QoS attributes in different UEC scenarios. For instance, a joint optimization approach is proposed in \cite{yu2020joint} under Things-UAV-Edge collaboration settings to optimize resource and task allocation. The proposed UEC scenario consists of edge servers, a UAV, and a few IoT devices. The UAV works as an edge server and relay in this study. The goal is to improve the latency for all IoT devices and the UAV's energy efficiency, {however, this study is still a single UAV scenario.} A similar study is carried out in \cite{liu2020joint} to reduce users' energy consumption by optimizing task offloading and resource allocation processes using the convex optimization method and stochastic learning automata algorithm. A set of UAVs is considered that work as edge servers instead of a single UAV. A cooperative UEC environment is studied in \cite{liu2020cooperative} where UAVs assist one another in executing computational tasks. The study aims to mitigate the interference from UAVs to devices while maximizing the long-term utility of the studied UEC network using deep reinforcement learning algorithms. It considers both centralized and distributed UEC environments. An age of information-aware resource management approach is proposed in \cite{chen2020age} where the freshness of information is considered during the maximization of each ground user's long-term computing performance. {The limitation of this study is its assumption assumes that the UAV and the ground users move at the same speed following a Markov mobility model, which is unlikely true in real-world scenarios.} The proposed approach formulate the problem as a stochastic game and proposes a novel online deep reinforcement learning scheme. A multi-UAV-based UEC environment is studied in \cite{seid2021multi} where a cluster of UAVs is provisioned to enable offloading services. The goal is to reduce the network's computation cost and meet the QoS requirements of IoT devices. It leverages a multi-agent deep reinforcement learning-based approach to reduce the cost. {However, all the above studies ignore the significant characteristic of UEC, the movement of UAVs.}

\subsubsection{\textcolor{black}{Joint Allocation and Provisioning}}

Joint resource allocation provisioning in UEC mostly involves optimizing UAV deployment, trajectory, and computing and communication resource allocation. For instance, an energy-efficient optimization approach is proposed in \cite{li2020energy} for resource allocation and trajectory planning. The study considers a scenario where a UAV acts as an edge server and collects and processes the computational tasks from users. The goal is to minimize the energy consumption of UAVs by jointly optimizing their trajectories, users' transmission, and computing task allocation. The study utilizes the following two algorithms to perform the optimization: a) the Dinkelbach algorithm, and b) the SCA algorithm. A three-layer online big data processing framework is proposed in \cite{wan2019toward} where a moving UAV collects and performs initial processing of raw data generated by IoT devices and sends it to the cloud. The aim is to efficiently allocate bandwidth and optimize path planning to efficiently collect distributed data from sensors. In this study, an online edge scheduling approach is leveraged to optimize the proposed resource allocation problem. The online path planning problem is solved using a deep reinforcement learning algorithm. An energy consumption minimization approach is proposed in \cite{jeong2017mobile} where a UAV offers computational offloading services to a set of mobile users. The approach aims at minimizing the energy consumption of mobile users while maintaining the expected QoS of the mobile application. An SCA-based algorithm is utilized to optimize energy consumption. The proposed approach considers both OMA and NOMA schemes for communication. {However, this is a single UAV scenario again. In addition, the authors assume that all the tasks can be purely completed by local resources.} In a similar setting, an energy consumption minimization approach is proposed in \cite{ji2020energy} where the aim is to improve the energy efficiency of the UAV and user devices. An ADM-based approach is proposed to optimize computing resources. Some studies consider a space-air-ground integrated network (SAGIN) to enable UEC. For instance, the authors in \cite{mao2020joint} utilize a SAGIN and assume that ubiquitous cloud services will be enabled via satellite. In this case, UAVs will provide computing services to remote IoT devices with the help of satellites. The aim is to minimize the service delay by optimizing task association, resource allocation, and deployment position using block coordinate descent and SCA-based approaches. {The shortage of this study is the ignorance of UAV trajectory in UEC.} A network function virtualization-based approach is leveraged in \cite{mei2019joint} to assist IoT devices in computational tasks. The study proposes a BCD-based algorithm to reduce the energy consumption of a UAV using optimal resource allocation and trajectory planning. {Unfortunately, the collaborations between UAVs or between UAVs and edge servers are not considered.}

\subsubsection{\textcolor{black}{Joint Provisioning, Offloading and Allocation }}

Several studies have considered all three aspects of resource management, including resource provisioning, computation offloading and resource allocation, to maximize the performance of different UEC systems \cite{zhao2021fairness,liu2021joint,liu2019uav,luo2021optimization}. A fairness-aware approach is proposed in \cite{zhao2021fairness} to optimally allocate resources and perform task scheduling. The goal is to minimize the energy consumption of a UAV where it acts as an edge server for multiple ground nodes. A mixed-integer nonlinear programming problem is formulated to solve the proposed optimization problem. The study utilizes an iterative algorithm to find the optimal solution considering the resource allocation, trajectory, bits scheduling, and task decision. The proposed approach considers fairness among the ground nodes and different operations of offloading services. A new type of UEC system is studied in \cite{liu2021joint} where edge computing services are provided via a UAV to a platoon of vehicles that are capable of wireless power transmission. The UAV may recharge from these vehicles to maximize its lifetime and computing power. An SCA-based method is leveraged to find the optimal solution. An energy-efficient scheduling problem is studied in \cite{dai2021towards} where multiple UAVs and base stations are provisioned to enable edge services. The main goal is to maximize the lifetime of mobile devices in the proposed UEC environment. It optimizes trajectories, task associations, and resource allocations to reduce energy consumption. The optimization problem has three sub-problems that are solved using a hybrid heuristic and learning-based scheduling algorithm. Collision avoidance is a critical research issue when dealing with multiple UAVs. A two-layer optimization approach is proposed to minimize the energy consumption of the ground users in \cite{luo2021optimization}. It jointly optimizes task scheduling, bit allocation, and trajectory. The first layer utilizes a bidding optimization method based on dynamic programming for the task scheduling process. The second layer leverages the ADMM to optimize the trajectory planning and bit allocation. It also considers a dynamic service strategy, optimal task scheduling, and a collision avoidance strategy.

\textcolor{black}{Mobility is a key concern in UEC environments as it may impact QoS significantly. Although most of the existing studies consider either IoT devices or UAV's positions to be fixed, there are a few studies that consider the mobility of both \cite{liu2019uav,ti2018joint}. For instance, a UEC environment is studied in \cite{ti2018joint} where both mobile users and UAVs may change their locations over time. The total power consumption is reduced using a joint optimization technique that considers task offloading, transmission power allocation, user association, and UAV path planning. The proposed problem is presented as a MINLP and solved with an iterative two-phase algorithm. A wireless-powered UEC environment is studied in \cite{liu2019uav} which assumes that a UAV is capable of energy transmission and provides IoT devices with energy and computation capabilities. The aim of this work is to minimize the UAV's energy consumption by finding the optimal offloading resource allocation, transmission power, bit allocation, and trajectory planning. This work claims that most of the studies do not take into account the effect of acceleration on energy consumption. Therefore, the effect of both UAV's velocity and acceleration are considered during the optimization. The study leverages SCA-based and alternative, decomposition, and iteration (DAI)-based algorithms to solve the problem.}


\section{Key Techniques and Performance Indicators}

Various resource management technologies are essential to achieve the optimal management of resources in UEC for achieving the QoS and QoE requirements of users. Recently, a number of state-of-the-art technologies have been proposed for resource management in UEC. According to the control of information exchange and strategy formulation in UEC systems, those technologies can be distinguished as centralized and distributed technologies. In general, centralized technologies mainly consist of convex optimization techniques, centralized machine learning, heuristic, greedy and iterative algorithms, while distributed technologies mainly consist of game theory, auction methods, blockchain and federated learning. In this section, we will discuss both centralized and distributed technologies and analyze their advantages and disadvantages. After that, we will summarize the main performance indicators used in the existing resource management studies in UEC, including energy consumption, throughput, latency, cost, and utility.

\begin{figure}
    \centering
    \includegraphics[width=.9\textwidth]{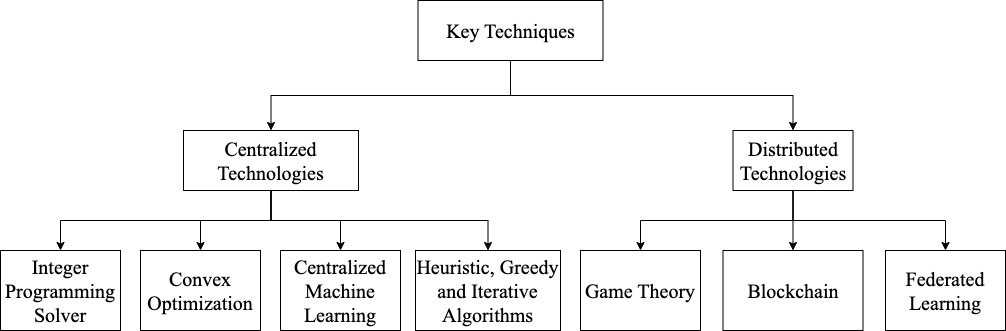}
    \caption{{A taxonomy of key techniques}}
    \label{fig:key_techniques}
\end{figure}



\subsection{Centralized Technologies}
Centralized technologies are widely implemented in the existing studies due to their advanced performance and intuitive design. In this section, we will review several widely-used centralized technologies.

\subsubsection{Integer Programming Solver}
\label{subsubsec:integer_programming_solver}

In most of the existing studies of resource management in UEC, including computation and data offloading, resource allocation and resource provisioning, the investigated problems usually are formulated as a 0-1 or mixed integer programming model \cite{zhang2021optimizing, mukherjee2020distributed, qu2021service, zhang2019computation, zhou2022two, deng2022uav}. Once the integer programming model is formulated according to the problem itself, including objective(s) and constraint(s), this mathematical model can be converted into equations as the input of integer programming solvers, such as Gurobi\footnote{http://www.gurobi.com/} and IBM CPLEX Optimizer\footnote{https://www.ibm.com/analytics/cplex-optimizer}. Since those solvers have been produced maturely for years, mathematical models can be implemented in different language versions of those solvers, including C, Java, Python and many others. For example, the integer programming technology is applied in \cite{guo2021coded} for finding the optimal solution to the mixed integer programming model formulated by coded distributed computing.

{
In general, integer programming solvers offer several advantages. Firstly, they are extensively developed and maintained by industry leaders like IBM. Secondly, this technology can be readily implemented for mathematical problem-solving, which is a commonly adopted approach in many existing studies on resource management in UEC. Thirdly, integer programming solvers are capable of finding optimal solutions to problems. Lastly, these solvers offer adaptability with various parameters, such as execution time.}
However, a significant drawback of this technology is its inherent high computational complexity. Solvers aim to find the optimal solution for given integer programming models, which typically involve NP-hard or non-convex problems. While certain solvers like CPLEX allow users to limit the execution time through parameters, setting strict time limits often leads to a substantial decrease in performance. For instance, in experiments conducted in \cite{xia2020cost, xia2022formulatinginter}, the performance of integer programming solvers was inferior to that of iterative approaches even when the time limit was set between 10 to 100 seconds.
Considering the low-latency service requirements in UEC, integer programming solvers are often impractical in real-world scenarios. Their limitations make them less suitable for most UEC applications.

\subsubsection{Convex Optimization}
\label{subsubsec:convex_optimization}
As mentioned in Section \ref{subsubsec:integer_programming_solver}, the problems investigated in resource management in UEC are typically NP-hard or non-convex, it is impractical to find the optimal solution by force-bruce algorithms or integer programming solvers. Since the convex optimization technology is mature in the area of wireless communication, many researchers attempt to convert a non-convex problem into a convex or near-convex problem, then apply a feasible convex optimization solution to solve the converted problem, such as \cite{su2021optimal}. Currently, the successive convex approximation (SCA), and Lyapunov optimization are decomposition techniques, the main techniques used in resource management for converting a non-convex problem into a convex or near-convex problem. For example, in \cite{tang2022uav}, the Lagrangian duality method and SCA technology are utilized for formulating computational offloading strategies to minimize energy consumption in UEC. The Lyapunov optimization technology is applied in \cite{lin2022novel} to dynamically allocate resources from UAVs to nearby mobile devices in UEC, aiming to complete computational tasks with minimum system cost and maximum system utility. Specifically, the authors formulate a queuing model for tasks on each edge server to keep the system stable. 


{When comparing the use of integer programming solvers with convex optimization techniques for resource management problems, it is evident that the computation complexity of the converted problems through convex optimization is significantly reduced. Additionally, convex optimization provides a theoretical bound that guarantees performance. However, it is important to note that the services offered by UEC require millisecond-level response times. Consequently, implementing convex optimization technology to solve resource management problems in UEC becomes challenging due to its relatively high computation complexity. Meeting the stringent time constraints while utilizing convex optimization techniques remains a significant obstacle in UEC resource management. In general, both integer programming and convex optimization are suitable for UAV and edge server deployment problems, since these problems typically do not have strict real-time constraints, allowing for the use of computationally intensive methods.}

\subsubsection{Centralized Machine Learning}
\label{subsubsec:centralized_machine_learning}
Recently, researchers are starting to solve resource management problems in UEC by machine learning, such as reinforcement learning, deep learning and others \cite{xue2022cost, zhao2022multi, liu2022deep, zhang2022deep, cheng2022deep, cheng2022learning, cheng2021intelligent}. For most techniques of integer programming solvers and convex optimization, they hardly handle the dynamics and uncertainty of UEC because they usually require complete information about everything, including edge servers, UAVs, devices, environment and mobility. Leveraging the benefits of machine learning, the dynamic resource management problems can be effectively solved by being formulated as Markov Decision process models, which can be solved by machine learning techniques, e.g., reinforcement learning. To name a few, Zhao et al. \cite{zhao2022multi} propose a deep reinforcement learning approach to solve the computation offloading problem in UEC. The proposed approach aims to achieve the minimum service delay and energy consumption, based on the Markov decision process. Similarly, Liu et al. \cite{liu2022deep} also implement a deep reinforcement learning approach based on the Markov decision process to solve the computation offloading problem in UEC with the only aim to minimize the total service latency.

Centralized machine learning technologies have their own advantages and disadvantages when tackling resource management problems. With a set of suitable parameters, machine learning technologies can provide near-optimal solutions to resource management problems in UEC, without complicated approach designs and high execution overhead due to the short time of prediction. {However, such technologies require significant computational resources during training with large datasets. The learning process itself can be time-consuming in order to achieve optimal performance. Furthermore, machine learning models often involve a vast number of parameters that require adjustment. 
Thus, centralized machine learning is more suitable for the algorithms running in the cloud server in a centralized manner to provide global control.
}

\subsubsection{Heuristic, Greedy and Iterative Algorithms}
Apart from the above technologies applied when solving resource management problems in UEC, the heuristic, greedy and iterative (HGI) algorithms are the most popular method to tackle those NP-hard or non-convex problems with short computational overheads. There is a special category of those algorithms, named approximation algorithms. The difference between approximation algorithms from other algorithms is that the approximation ratio, i.e., the ratio of performance of the optimal solution over the performance of the solution provided by the approximation algorithm, can be theoretically analyzed. In this way, the theoretical performance can be guaranteed. For most of HGI algorithms, they are normally designed from the view of the optimization objective, associating with the existing techniques such as greedy algorithms, dynamic programming and local search and relaxation \cite{lee2022uav, zhang2022joint, yu2022uav, chiu2022collaborative, tao2021uav}. In \cite{hou2019fog}, the authors propose a heuristic algorithm for formulating computation offloading strategies, aiming to minimize energy consumption. Chen et al. \cite{chen2022joint} implement a greedy algorithm with genetic algorithm operators to solve the joint computation offloading and deployment optimization problem in UEC, aiming to minimize the task response time. 

Compared with the methods in Sections \ref{subsubsec:integer_programming_solver}, \ref{subsubsec:convex_optimization} and \ref{subsubsec:centralized_machine_learning}, HGI algorithms can solve the NP-hard or non-convex problems with low computation complexity. Those approaches are relatively simple, direct and easy to be implemented. Especially for approximation algorithms, they can also provide a theoretical bound to ensure performance in the worst cases. However, as the price of a relatively simple design, HGI algorithms may fall into the local optimal trap, weakening the performance on achieving optimization goals of resource management in UEC. {Benefiting from high efficiency, HCI algorithms are more suitable for real-time decision-making scenarios in UEC.}

\subsection{Distributed Methods}

In UEC, edge servers and UAVs are geographically distributed to cover different areas wirelessly. Considering the nature of distributed computing paradigm, a central control applied in centralized methods can become a potential bottleneck from perspectives of resource, location and security. In this section, we will review several widely-used distributed technologies.

\subsubsection{Game Theory}
As a powerful tool, game theory has been widely applied for solving edge computing problems in a parallel and decentralized manner \cite{he2019game, he2021game, cui2020interference, cui2021demand, ning2020mobile, shen2023game}. Under the game theory framework, the iterations among multiple players, such as UAVs and devices, can be analyzed to devise incentive and compatible mechanisms for finding collectively satisfactory solutions. The action of each player can be changed based on other's actions for maximizing their interests. Once a Nash Equilibrium achieves, no further actions will be changed and the solution is formulated. For example, the authors of \cite{han2022age} investigate the age of information minimization problem and propose a stochastic game theoretical approach. In addition, they propose a learning-based algorithm to achieve the Nash Equilibrium. Tackling a similar research problem, Wang et al. \cite{wang2022data} formulate it as an ordinary potential game to minimize the age of information. 


{Game theoretical approaches, as distributed technologies, offer the advantage of parallel execution, which significantly reduces the overhead associated with formulating solutions. This reduction in execution overheads leads to a decrease in service latency, making game theoretical approaches well-suited for real-time decision-making scenarios.}
In game theory, reaching a Nash Equilibrium is crucial for formulating solutions. However, it's important to note that there can be multiple Nash Equilibria in a game. The performance of a game theoretical approach heavily depends on the selection of the Nash Equilibrium for solution formulation. Obtaining the Nash Equilibrium with optimal performance poses a challenge in game theoretical approaches.
Furthermore, game theory involves continuous iterations of information exchanges and decision updates to achieve the Nash Equilibrium. These information exchanges may introduce high communication overhead, which may impact the overall efficiency of the approach. Managing and minimizing this communication overhead is essential for ensuring the effectiveness and scalability of game theoretical solutions.

\subsubsection{Blockchain}
As a decentralized system, blockchain has attracted significant attention from academia and industry because of its distributed, irreversible and traceable functions, etc. Blockchain has been applied in many applications in edge computing, including data sharing \cite{yuan2021csedge, abdellatif2021medge}, resource allocation \cite{zhang2021resource}, intrusion detection \cite{liu2021blockchain}, identity authentication \cite{wang2021blockchain}, etc. Recently, some blockchain systems have been implemented for supporting resource management in UEC, including \cite{nilsen2022competing, li2022solving, latif2022sdblockedge, singh2022derived}. To name a few, the authors of \cite{masuduzzaman2022uav} propose a blockchain-based approach to manage automated traffic in UEC, with a vehicle detection algorithm. With blockchain, the automated traffic management scheme can be ensured securely. In \cite{wang2022airbc}, the authors design a lightweight reputation-based blockchain scheme to improve both security and efficiency when managing resources in UEC, with an improved Practical Byzantine Fault Tolerance consensus mechanism.

Compared with other distributed technologies, blockchain is more advanced in data integrity and security, due to the irreversible and traceable characteristics of blockchain. 
{Given the distributed nature of UEC and the resource limitations of edge servers and UAVs, ensuring security and privacy becomes a more complex task. These challenges can be addressed through the use of blockchain technology. Therefore, blockchain is a viable solution when there is a need to safeguard the security and privacy of servers, users, and data.} However, blockchain also has its disadvantages. Firstly, the latency caused by generating transaction blocks is usually high, impacting the overall performance of blockchain-based approaches. Secondly, the incentive mechanism among edge servers, UAVs and devices should be carefully considered for invoking them actively during resource management in UEC. In addition, the resources on edge servers, UAVs and devices are limited, however, the resources required for hosting blockchain are significant. Thus, lightweight blockchain solutions are required for ensuring the performance and security of resource management in UEC.

\subsubsection{Federated Learning}
As a prospective machine learning scheme, federated learning has emerged to address the privacy and performance issues of traditional centralized machine learning approaches \cite{li2020federated}. By federated learning, data collected by devices can be hidden from edge servers and UAVs since only the training metrics are uploaded from devices. In the past years, federated learning has been applied to resource management in UEC \cite{cheng2022auction, cheng2021joint, zhang2020fenghuolun}. For example, the authors of \cite{rahbari2021fast} introduce a federated learning approach to minimize the latency and energy consumption during computational tasks and data offloading among UAVs. In this study, a swarm of UAVs share resources and collaborate to improve the efficiency of communication and offloading. Nie et al. \cite{nie2021semi} aim to minimize the consumed power during the communication and computation process in UEC. To deal with the user association, resource allocation and power control problems, the authors propose a multi-agent federated learning approach in a semi-distributed manner.

As mentioned above, the privacy of users' data can be protected by federated learning, since the raw data is only stored on the device, UAV or edge server where such data are collected. This also releases the transmission burden in wireless networks. In addition, federated learning helps collaborate resources from devices, UAVs and edge servers to complete the highly complex training process, which cannot be completed by a single device. However, federated learning also faces shortages. The existing federated learning-based approaches for resource management in UEC face malicious attacks, e.g., data poisoning attacks \cite{tolpegin2020data}. Under the attacks, the performance of those approaches cannot be guaranteed. {In conclusion, federated learning is a good option for processing learning tasks in UEC with limited computation and storage resources.}.

\subsection{Key Performance Indicators}

\textcolor{black}{
In this section, we discuss the key performance indicators, i.e., energy consumption, throughput, latency, resource utilization, utility and cost in UEC for resource management. These indicators are identified based on the objectives presented in table \ref{tab:resource_prov} - \ref{tab:joint_resource}.}

\subsubsection{Energy Consumption}
\textcolor{black}{
A key challenge of resource management in UEC is to deal with limited energy. Both IoT devices and UAVs have limited power. Many studies in UEC, therefore, focus on minimizing the energy consumption of either IoT devices, UAVs, or both \cite{hu2018joint,liu2020joint, dai2021towards, hu2019task, ti2018joint, luo2021optimization}. The existing studies mainly consider two aspects of energy consumption in UEC: a) computation and transmission energy consumption of IoT devices, and b) computation, transmission, flying, and hovering energy consumption of UAVs. The energy consumption for computation is typically measured by the required number of CPU cycles for computational tasks. The transmission energy consumption considers both offloading and receiving the related energy consumption. The transmission energy consumption depends on the used communication protocols (i.e., OMA or NOMA). Therefore, several studies proposed energy consumption minimization approaches for OMA and NOMA separately \cite{ji2020energy,jeong2017mobile,li2020energy}. Many studies often ignore the energy consumption caused by hovering or acceleration or assume that the UAVs have fixed speed and height \cite{liu2020joint, yu2020joint, ti2018joint, wan2019toward}. However, the speed, acceleration, hovering, and height of UAVs considerably impact energy consumption. The energy consumption for hovering or acceleration is considered in several studies \cite{zhao2021fairness, jeong2017mobile, ji2020energy}. Energy consumption is often optimized jointly with other QoS such as latency, throughput, and resource utilization \cite{ning2021dynamic,yu2020joint,xu2020big}}.   


\subsubsection{Throughput}

\textcolor{black}{
Throughput is one of the most important performance indicators in UEC to evaluate and compare various resource management approaches. Therefore, many existing studies have intensively studied throughput in various UEC scenarios \cite{wang2019online, cheng2018uav, he2021multi, wu2020cell}. Most of the existing studies mainly focus on either computational throughput or network throughput or both. Several studies focus on maximizing each IoT device's computational capacity \cite{dai2020energy, chen2020age}. For instance, an age of information-aware resource management approach is introduced in \cite{chen2020age}, where the aim is to maximize each mobile user's long-term computation performance using a novel online deep reinforcement learning scheme. Several studies focus on maximizing the computational throughput of all IoT devices \cite{hu2020wireless, wu2020cell, cheng2018uav}. For instance, a cooperative offloading and resource management approach is proposed in \cite{liu2021joint} where the aim is to maximize the system-wide computation capacity. Several studies also focus on improving the throughput of the UAV in a UEC system \cite{zhou2018computation}. Throughput is also jointly considered with other performance indicators such as resource utilization, cost, utility, and energy efficiency \cite{feng2021hybrid}.}  


\subsubsection{Latency}

\textcolor{black}{
Latency is another key performance indicator directly impacting the user experience in UEC. For delay-sensitive applications, latency is a critical QoS attribute. Many studies focus on improving the latency in UEC \cite{sun2020optimizing, mao2020joint, wang2021computation}. Latency primarily depends on three key factors which are local computation time, transmission time, and remote computation time. Local computation time refers to the required computation time by IoT devices or UAVs (as users). The transmission time refers to the data sending time and the result receiving time. The remote computation time refers to the required time for processing offloaded data. Improving the latency in UEC systems is considered a complex optimization problem. Most of the existing studies focus on specific scenarios to develop various approaches to improve latency along with other QoS \cite{yu2020joint, mao2020joint}.}   



\subsubsection{Cost} 

\textcolor{black}{
Several studies aim at minimizing the cost of resource management in UEC systems \cite{zhang2020energy, guo2021coded, seid2021multi}. The cost in UEC is typically defined in terms of multiple cost factors in resource management such as offloading or transmission cost, energy cost, and processing cost. Hence, the cost is considered as a compound performance indicator in UEC. Most of the studies typically develop different types of cost models targeting specific application scenarios \cite{wang2017resource,asheralieva2019distributed,seid2021multi}. For instance, A cost model for mobile devices is proposed in \cite{zhang2020energy} where cost is considered as a linear combination of the energy consumption of the mobile devices and the corresponding time latency. The proposed cost model is applied to a scenario where each mobile device can execute its tasks locally or offload tasks to UAVs or Base stations. The total system cost is minimized in \cite{guo2021coded} where the total system cost is defined as the sum of the UAVs costs in whole task completion.} 


\subsubsection{Utility} 
\textcolor{black}{
The utility in a UEC system typically indicates the user satisfaction achieved under a particular resource management strategy. The existing work usually measures the utility using pre-defined utility functions. A utility function is represented by different QoS attributes such as latency, throughput, energy consumption, data rate and cost. Utility functions often require mathematical transformation as each QoS attribute may have a unique unit of measurement. The transformation is typically performed using reciprocal, logarithm, and weighted summation methods. The utility function is then maximized using various resource management algorithms to achieve expected user satisfaction \cite{apostolopoulos2021data,liwang2021let, khochare2021heuristic, liu2020cooperative}.}


\subsubsection{Others} 

\textcolor{black}{
Some of the existing studies consider resource utilization, profit, and UAV coverage as performance indicators of their proposed resource management strategies. For instance, a resource pricing and trading scheme are proposed for the edge server's profit and UAV's revenue. A novel path planning and resource management strategy is proposed for big data processing in UEC \cite{wan2019toward} where the aim is to enhance service area coverage. Resource utilization is another key concern in UEC due to the limited available resources. Therefore, several studies focus on maximizing network, computation, and storage resource utilization. Resource utilization is typically measured by the ratio of the utilized resources and the total available resources. Resource utilization is typically studied with other QoS attributes such as energy consumption, latency, and throughput.}

\section{Research Challenges and Directions}

The research on resource management in UEC has accumulated a number of results. However, many challenges and research gaps in this area still have not been well studied. In this section, several existing challenges and future research directions are discussed.

\subsection{Architecture}

The three-layer architecture of a UEC system typically consists of a things layer, an edge layer, and a cloud layer. The architecture shown in Fig. \ref{fig:architecture} captures the deployment aspect of physical resources in UEC. However, most of the existing studies only consider a partial architecture when performing efficient resource management in UEC, which ignores one or several of the collaborations among Things, UAV, edge servers and cloud, as shown in Table \ref{tab:collab}. Such a partial architecture weakens the feasibility and functionality of UEC systems, and results in resource waste. This violates the original pursuit of resource management in UEC. In this case, the researchers should consider the architecture more comprehensively to manage the resources of UEC systems in a more effective way. 

Moreover, the deployment of software components in different layers is not investigated in the existing literature. It may be because of the fact that most of the existing studies are focusing on theatrical aspects of resource management in UEC. In the real-world scenario, resource deployment may face many technical challenges that could be overcome by innovative architectural patterns for the deployment of software components \cite{tran2020anatomy}. In such cases, future research will need to investigate various aspects of resource management on virtual levels such as virtual machine migration and consolidation, serverless computing, and auto-scaling. Another important aspect is the integration of space-based networks in the architecture. In the near future, space, air and ground network (SAGIN) will be widely deployed to support billions of IoT devices. In this case, heterogeneous networks and resources need to be considered for efficient resource management. 

\subsection{UAV trajectory design} 
As a significant research problem in UEC, UAV trajectory design is key to the success of resource management studies in UEC. Among the existing studies, a small number of researchers consider the UAV trajectory design when they tackle resource allocation and computation offloading problems in UEC, such as \cite{li2020energy, miao2022drone}. The mobility of UAVs is one of the important characteristics of UEC, while the trajectory of UAVs can result in various changes in user coverage, wireless interference and energy harvesting. Without proper UAV trajectory designs, the feasibility and usability of resource management approaches in UEC cannot be guaranteed. For the existing UAV trajectory studies, the proposed approaches are mainly designed for a single UAV and ignore the conflict of multiple UAV flight paths. In most real-world UEC scenarios, a set of UAVs are involved to provide high-performance and low-latency services for users, e.g., mobile and IoT devices. From the view of feasibility and usability, the multiple-UAV trajectory design with environmental changes should be considered for investigating resource management problems in UEC. Otherwise, the gap from research to industry cannot be filled.

\subsection{Energy charging and harvesting}
Considering the fact that UAVs and IoT devices have constrained battery life, saving energy consumption and maintaining the long life of installed batteries are important when providing high-performance and low-latency services in UEC. Moreover, from a service provider's perspective, a long-term service goal is normally more significant than a short-term service goal \cite{xia2021online}. To provide long-term and high-quality services, an effective method for tackling such a challenge is to implement wireless charging or energy harvesting techniques in UEC. The existing studies with consideration of energy harvesting mainly implement wireless charging and energy harvesting either from edge servers to UAVs or from UAVs from IoT devices \cite{zhou2018computation, feng2021hybrid, wang2017resource, xu2018uav}. In addition, those studies assume that such energy transfer is stable and continuous, which is not always true in real-world UEC scenarios. Due to many external factors, such as wind, rain and interference, the energy transfer process could be easily impacted, even interrupted. In this case, the design of resource management strategies in UEC becomes very challenging. Thus, the energy transfer process needs more attention from researchers when they investigate resource management problems in UEC with consideration of wireless charging and energy harvesting. Besides the energy transfer, another way for energy harvesting is utilizing green energy, such as light and wind, to support UAVs and IoT devices. With the assistance of green energy, the provided services can be more stable which would help reduce environmental pollution.

\subsection{Security and Privacy}
\textcolor{black}{Security and privacy are two major concerns in any system. Managing security and privacy is more challenging in UEC due to the distributed architecture, limited resources, and mobile and volatile environment \cite{ouyang2021trust,zhou2019secure, patrikar2022anomaly}. However, the existing literature has paid little attention to the security and privacy aspects of resource management in UEC. A failure of UAVs caused by attacks or system failures may affect the reliability and robustness of an entire UEC system \cite{gao2020securing}. Therefore, the resource management research in UEC needs to investigate different types of attacks and fault tolerance strategies for UAVs. The UEC systems heavily rely on communication channels and protocols between various components. Information could be compromised during the data transmission \cite{yao2021secure}. The enabling resource-intensive security protocols may impact the performance of UEC systems. Therefore, various lightweight security protocols need to be investigated for resource management in UEC. Most of the existing studies for resource management unconditionally trust the user devices. However, the trust of user devices needs to be considered while accepting offloading requests in UEC. If a user's device is compromised or acts as a malicious node, UAVs require an efficient and fast way to identify it. Moreover, privacy-sensitive data may need to be processed in UEC depending on various application scenarios such as smart health or military operation. In such cases, privacy-sensitive data processing strategies need to be considered. Integration of privacy-enabled algorithms may impact the performance of the UEC systems. Therefore, effective trade-off mechanisms between performance and privacy are required for resource management.}

\subsection{External Factors}
\textcolor{black}{The environmental factors such as wind, fog, rain, and storm may significantly affect the performance of UAVs \cite{alkouz2020swarm}. However, most of the existing studies for resource management in UEC ignore the environmental factors for resource management. The trajectory, speed, and acceleration of a UAV are required to be controlled effectively in various weather conditions to optimize their energy consumption. Another key concern in UEC resource management is the interference in multi-UAV systems. When multiple UAVs communicate in a shared channel with user devices, interference may arise \cite{lu2020resource}. Therefore, effective communication models and trajectories need to be designed efficiently to avoid interference. Most of the existing studies assumed that all UAVs are operated by the same provider. However, in the near future, the same location may get coverage by multiple UAVs operated by different providers. In such a case, the trajectories of UAVs need to be designed carefully to avoid collision. Recharging on UAVs is another key concern that is overlooked in most of the existing studies, which also assumes that a UAV has a fixed amount of energy that needs to be optimized to deliver computing services. However, A UAV may have an opportunity to recharge its battery from charging stations or via wireless energy transfers \cite{alkouz2021reinforcement}. Therefore, the computational performance of the UEC systems can be improved by considering available charging options for UAVs while designing resource allocation approaches in UEC. Another key challenge of UEC is to comply with government regulations. Most of the studies in UEC do not take into account government regulations. In some countries, governments may impose specific restrictions on the maximum and minimum height of UAVs. The height of UAVs affects the communication between IoT devices and UAVs in the UEC systems. Therefore, future research needs to consider government rules and regulations during UAV deployment and trajectory design in UEC.}

\subsection{Real-world experiments}
Currently, the evaluations of resource management strategies are generally based on simulation technologies rather than any real-world implementations. Most of the existing studies evaluate their approaches in simulated scenarios by Java, Python, Matlab and other programming languages with real or synthetic data sets, such as \cite{xu2022efficient, lee2022uav, tan2022joint, zhu2022blockchain, mahmood2022distributed}. Assumptions on weather, power, speed, etc. are needed to enable the system simulation. However, the connection to real-world systems and the feasibility of the proposed strategies cannot be convincingly demonstrated in this case. Thus, without thorough test systems in real-world scenarios, the evaluations are not comprehensive and less confident. From this point, more efforts are required to evaluate the feasibility of proposed approaches in real-world systems. In addition, real-world or highly simulated test beds are in high demand.

\section{Conclusion}

Efficient Resource management is a key requirement for deploying UAV-enabled Edge Computing. We present a comprehensive survey on the state-of-art of resource management in UEC. The resource management in UEC is considered from resource provisioning, computational task offloading, resource allocations, and joint resource management perspectives. We discuss the key techniques and performance indicators for efficient resource management in UEC. A conceptual architecture is presented that illustrates the collaborations and communications between UAVs, IoT devices, edge servers, and cloud servers. Existing study shows that UAV-enabled edge computing is a promising technology that may facilitate a wide range of innovative applications in various domain such as urban traffic control, smart agriculture, disaster response, tactical edge, and logistics. Although most of the existing studies focus on building a theoretical foundation for efficient resource management in various scenarios, the wider adoption of UEC technology is still hindered by several key challenges. For instance, architectural design decisions in the environment may considerably impact the quality attributes of UEC-based applications. However, it remained largely unexplored in the state-of-the-art. External factors such as environment, interference, recharging availability, and government regulations must be considered before deploying UEC technologies. The operating environment of UEC is intrinsically volatile, resource and energy-constrained. Therefore, maintaining security and privacy in UEC systems is critical and requires more intensive research. A major challenge in conducting research in UEC is the lack of real-world testbeds for conducting experiments. Building real-world testbeds or UEC simulation platforms that re-assemble the real world will attract more research interests in this domain. Another future research direction can be found by considering the interests of different stakeholders in UEC. Most of the existing studies assume that one stakeholder is responsible for managing all resources in IoT devices, UAVs, edge servers, and cloud servers. In the real world, these resources may be provisioned by different stakeholders who will provision their resources based on their economic models. In such scenarios, the resource management approaches would require careful consideration of the economic models of different stakeholders. For example, some stakeholders may deploy charging stations in remote areas where UAVs may recharge themselves and continue to provide computing services. Another group of stakeholders may provision UAV-as-a-Service where UAVs can be rented for a period of time. In such scenarios, managing resources efficiently could be more challenging, given the constraints imposed by different stakeholders.

\bibliographystyle{ACM-Reference-Format}
\bibliography{Main}

\end{document}